\documentclass[aps,pra,a4paper,10pt,twocolumn,nofootinbib]{revtex4-2} 

\usepackage[T1]{fontenc}
\usepackage{mathptmx}
\usepackage{datetime}
\usepackage{amsmath}
\usepackage{amsfonts}
\usepackage{amssymb}
\usepackage{bbding}
\usepackage{eufrak}
\usepackage{mathrsfs}
\usepackage[mathscr]{euscript}
\usepackage{fancyhdr}
\usepackage{colordvi}
\usepackage{hyperref} 
\usepackage{epsfig}
\usepackage{color}
\usepackage{bm}
\usepackage{enumerate} 

\def\overstrike#1#2{{\setbox0\hbox{$#2$}\hbox to \wd0{\hss
    $#1$\hss}\kern-\wd0\box0}}

\def\cRed#1{{{#1}}}

\newdateformat{yymmdddate}{\THEYEAR/\twodigit{\THEMONTH}/\twodigit{\THEDAY}}

\def\APPENDSUP{the Supplementary Material}

\def\WdT{W_{\textup{dT}}}
\def\WdTs{W_{\textup{dT}}^{*}}
\def\LdT{L_{\textup{dT}}}

\def\KPI{K_\textup{P}}
\def\API{A_\textup{P}}
\def\DST{Dst}

\def\TEC{\tau_{\textrm{EC}}}
\def\dTEC{\delta_{\textrm{EC}}}

\begin{document}
\title{$\LdT$: An indicator of ionospheric activity
 based on
 statistical distributions
 in GNSS-derived TEC rates of change}

\author{Paul Kinsler}
\homepage[]{https://orcid.org/0000-0001-5744-8146}
\email{Dr.Paul.Kinsler@physics.org}
\author{Biagio Forte}
\homepage[]{https://orcid.org/0000-0003-1682-1930}
\affiliation{
  Department of Electronic and Electrical Engineering
  University of Bath, Bath BA2 7AY,
  United Kingdom
}

\lhead{\includegraphics[height=5mm,angle=0]{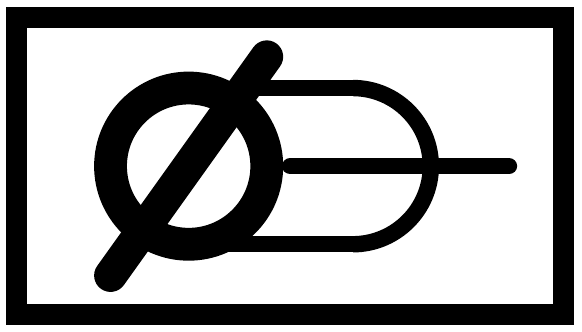}~~PAGIONO}
\chead{$\LdT$: A GNSS ionospheric activity index}
\rhead{
\href{mailto:Dr.Paul.Kinsler@physics.org}{Dr.Paul.Kinsler@physics.org}
}

\begin{abstract}

Many aspects of our societies now depend upon satellite telecommunications, such as those requiring Global Navigation Satellite Systems (GNSS). GNSS is based on radio waves that propagate through the ionosphere and experience complicated propagation effects caused by inhomogeneities in its electron density. The Earth's ionosphere forms part of the solar-terrestrial environment, and its state is determined by the spatial distribution and temporal evolution of its electron density. It varies in response to the ``space weather'' combination of solar activity and geomagnetic conditions. Notably, the radio waves used in satellite telecommunications suffer due to the dispersive nature of the ionospheric plasma.

Scales and indices that summarise the state of the solar-terrestrial environment due to solar activity and geomagnetic conditions already exist. However, the response of the ionosphere to active geomagnetic conditions, its geoeffectiveness, and its likely impact on systems and services are not encapsulated by these. This is due to the ionosphere's intrinsic day-to-day variability, persistent seasonal patterns, and because radio wave measurements of the ionosphere depend upon many factors.
Here we develop a novel index ($\LdT$) that describes the state of the ionosphere \cRed{-- as is relevant to GNSS --}  during specific space weather conditions. It is based on propagation disturbances in GNSS signals, and is able to characterise the spatio-temporal evolution of ionospheric disturbances in near real time. This new scale encapsulates day-to-day variability, seasonal patterns, and the geo-effective response of the ionosphere to disturbed space weather conditions; and can be applied to data from any GNSS network. It is intended that this new scale will be utilised by agencies providing space weather services, as well as by service operators to appreciate the current conditions in the ionosphere, thus informing their operations.

\end{abstract}


\date{\today}
\maketitle
\thispagestyle{fancy}

%
\section{Introduction}\label{S-context}

The Earth's ionosphere is part of the solar-terrestrial environment,
 and responds to space weather conditions in synergy
 with the plasmasphere and the magnetosphere \cite{Kelley-EIONO}.
The more extreme responses typically 
 originate in solar events such as flares and coronal mass ejections, 
 and have an immediate consequence for application and services
 reliant upon radio signals that propagate through the ionosphere.
Because the ionosphere is a plasma, 
 its particular state affects the propagation of radio waves, 
 notably those with frequencies between VLF and C band.
In particular, 
 instabilities \cite{Kelley-EIONO}
 and irregularities \cite{Aarons-1982ieee,Basu-MB-1988rsc,Forte-CSHMDPKB-2017jgrsp,Cherniak-KZ-2014rsc,Zakharenkova-CK-2019jgrsp,Cherniak-Z-2016eps,Skone-2001jg,Moraes-VCAPSMSFS-2018gs}
 in the ionospheric plasma
 can have practical and end-user effects on GNSS positioning -- 
 notably increased error, 
 or in more extreme cases,
 a loss of (usable) GNSS signals.
In this article,
 we present a logarithmic-scale index ($\LdT$) that 
 can be applied on wide range of spatial and temporal scales, 
 and encapsulates this ionospheric variability
 based on how the ionosphere affects the GNSS radio signals
 passing through it; 
 thus assisting systems and services in adopting suitable countermeasures
 that limit the impact 
 \cite{Forte-AAAVMSKJ-2024asr,Ishii-BFHBR-2024asr}.

Of course, 
 there are already a range of geomagnetic indices currently in widespread use, 
 but none of these have a direct relationship to the ionospheric state.
The most notable is $\KPI$ \cite{Matzka-SYBM-2021sw}, 
 which is based on combined magnetometer measurements
 from a specific set of ground locations; 
 $\API$ is a daily averaged version of $K_p$; 
 and then
 $\DST$ \cite{Sugiura-1964aigy,Nose-ISK-2015} is based on an estimate of 
 the globally symmetrical equatorial electrojet
 (``ring current'')
 in the magnetosphere.
Whilst these indicate the presence of geomagnetic disturbances
 capable of modifying the global state of the ionosphere,
 they are not sensitive to geoeffectiveness
 in the ionospheric response to active geomagnetic conditions. 
Moreover,
 these indices are not easily related to ionospheric propagation effects 
 that can impact \cRed{GNSS} services,
 or other operations
 reliant upon satellite telecommunications.

Although the ionosphere can be probed directly 
 by using ionosondes,
 incoherent scatter radars
 or
 special purpose satellites, 
 these
 only provide limited coverage.
However,  
 since the plasma density and its variability can be inferred using 
 observations of its propagation effects on radio signals, 
 the state of the ionospheric plasma can be appreciated in near-real time
 on a global scale by using GNSS, 
 allowing the creation of 
 maps of the Total Electron Content (TEC)
 and 
 measures derived from its rate of change, such as ROTI
 \cite{Pi-MLH-1997grl,Jakowski-SSK-2006asr,Gulyaeva-S-2008ag,Jakowski-BW-2012rsc,
Cherniak-KZ-2014rsc,Cherniak-Z-2016eps,Cherniak-KZ-2018eoi,Wilken-KJB-2018jswsc,
Jakowski-H-2019sw,Denardini-PBNCRMCRSB-2020sw,
Kotulak-ZKCWF-2020rse,Kotulak-KFFWLB-2021sen,John-FAAAVHS-2021rsc,
Fabbro-JLR-2021jswsc,Kotulak-KFFWLB-2021sen}.
These maps are estimated from a global network of GNSS monitoring stations
 and are provided by the International GNSS Service, 
 although these are also cross-validated 
 by a range of approaches provided by different Institutions
 \cite{HernandezPajares-JSB-2002asr,Juan-SRGIP-2018jswsc,Bruyninx-LFP-2019gpss,Borries-WJGDKJTHFH-2020asr}.
Such
 TEC maps provide insights on the overall distribution of ionisation
 in response to current solar and geomagnetic conditions, 
 whereas maps based on ROTI --
 or other similar indicators \cite{Nykiel-CHJ-2024gpss} --
 provide an indication of irregularities forming in different regions.
This ``irregularity detection'' occurs since 
 irregularities along a given ray propagation path introduce 
 temporal fluctuations in the received phase and intensity of GNSS radio waves.

However, 
 a limitation in the current approach to ROTI maps
 lies in their consideration of an average ROTI value in 
 predefined map pixels.
The typical discrete pixelization of the globe over
 a hypothetical screen at F-region altitudes (e.g., 350 km)
 therefore covers an area much greater than the size of any individual disturbance, 
 so that the averaging greatly reduces the sensitivity
 to geometry-dependent propagation disturbances, 
 and hence to the presence of irregularities in any given pixel. 
However, 
 ROTI is generally computed for specific latitude and longitude pixels,
 in order to create time-dependent maps of ionospheric disturbances; 
 and there is as yet no widely agreed mechanism to merge these 
 spatially localised results into a single regional or global number, 
 although various proposals exist
 (see e.g. \cite{Wilken-KJB-2018jswsc,Liu-HLNYMGBWOO-2021sw,Juan-SRGIP-2018jswsc}.
Further, 
 various approaches to improving the sensitivity to disturbances,
 and the detection of irregularities
 have been suggested:
 for example,  
 the GIX \cite{Jakowski-H-2019sw,Nykiel-CHJ-2024gpss},
 a probabilistic description \cite{Forte-AAAVMSKJ-2024asr}
 as well as others \cite{Wanninger-2004iongnss,
Pignalberi-PRG-2018sge,Roberts-FJ-2019gpss,Koulouri-SVRAF-2020jg,Liu-HLNYMGBWOO-2021sw,
John-FAAAVH-2021rse,Liu-HLNYMGBWOO-2021sw,Flisek-FFKKplus-2023jswc}.

In this work we introduce
 a complementary method for estimating the spatial and temporal variability
 of propagation disturbances \cRed{that can affect GNSS}, 
 but which -- 
 like ROTI --
 relies on the 
 temporal rate of change of TEC (dTEC) evaluated over the IGS network; 
 \cRed{but differs from the scintillation based measures $S_4$ or $\sigma_{\phi}$
 \cite{Forte-AAAVMSKJ-2024asr,Roberts-FJ-2019gpss,Koulouri-SVRAF-2020jg}}.
In Section \ref{S-ourindex} we introduce the 
 main concepts behind our index.
Then, 
 in Section \ref{S-methods} we describe how we analyse and partition the 
 GNSS data to create histograms of dTEC values; 
 followed by Section \ref{S-fitting}
 where we reduce these histograms to a single ``width'' parameter $\WdT$
 by fitting them to a weighted set of functions.
In Section \ref{S-dTECnormalisation}
 we explain our normalisation process, 
 a method which enables us to
 treat dTEC values obtained at different elevation angles 
 and from different band pairs on the same footing.
As an example of how repartitioning a set of data 
 can be useful, 
 in Section \ref{S-g2e-utcvslocal} we compare dTEC values --
 and their resulting fitted widths -- 
 divided by local time as well as UTC; 
 the local time partioning enhances the visibility of diurnal activity
 in the ionosphere.
Then, 
 in Section \ref{S-L-index}
 we define our $\LdT$ index and apply it 
 to a range of geomagnetic conditions, 
 notably two very quiet days in January 2022, 
 as well as storm periods (April 2023, May 2024).
Finally, 
 in Section \ref{S-conclusion} we conclude.

\section{Preview}\label{S-ourindex}

\cRed{Here we are focussed on demonstrating how GNSS-based
 TEC inferences from disparate constellations,
 sky-locations, 
 and band frequencies
 can be unified in a systematic manner.}
Our
 $\LdT$ index therefore focusses
 on the statistical properties
 of temporal changes in TEC (``dTEC'') as an indicator in its own right.
By first characterising a 
 reference data set representing the minimal or ``quiet'' behaviour 
 of the ionosphere --
 as characterised by GNSS signals --
 we can then appreciate the conditions implied
 during more active ionospheric periods.
Further, 
 by analysing and normalising signal data from all GNSS band pairs,
 as divided into both temporal and spatial intervals, 
 we can investigate how these affect the assessment and reporting
 of the ionospheric state.
For this purpose 
 we use ``dTEC'' as an abbreviation for the 
 change in TEC per second, 
 as computed by tracking signals between each satellite- ground station pair, 
 and based on 30 second interval data.

A key point is that although it might be tempting to 
 characterise the dTEC distributions obtained via GNSS
 using measures such as average, 
 variance, 
 kurtosis,
 and higher order moments;
 we see in the data that 
 the distributions contain power-law tails 
 that become visible at large dTEC values.
In some cases, 
 the power exponent indicating the tail's fall-off is 
 low enough so that some of these moments become infinite.
So although one can compute any moment from the sampled data, 
 a reliable relationship between that computation 
 and that of the underlying distribution is not always guaranteed.

Accordingly, 
 instead of computing moments we instead match
 a set of fitting functions
 to the distribution based on a frequency histogram
 of the set of dTEC values.
These cannot always fit every one of the dTEC distributions we see
 to very high accuracy, 
 but they encapsulate most typical features, 
 and give us access to estimations for true width properties  
 of the distributions.
This finite width measure can be straightforwardly converted into an index, 
 whilst the full set of fitting parameters
 can be used to monitor the variation in distribution properties
 with ionospheric conditions.

In contrast to ROTI and its variants, 
 our index has a number of advantageous features, 
 as we will show in more detail below.
Firstly, 
 and in contrast to current applications of ROTI, 
 $\LdT$ is based on a statistical or distributional approach, 
 and so without additional assumptions
 it can be applied equally well to any dataset of dTEC values
 (e.g., smaller regional networks, or denser global networks) --
 however they might be divided or localised in time, 
 space, 
 or by other criteria.
Secondly, 
 since ROTI is based on a computation of variance, 
 it has a specific technical shortcoming --
 we will see below that distributions
 of dTEC values have power-law tails, 
 and therefore are not always guaranteed to have finite variances.

%
\section{Methods \& Algorithms}\label{S-methods}

The $\LdT$ index that we will introduce later in Section \ref{S-L-index}
 is derived from a
 characterisation measure for ionospheric disturbances, 
 and that is based on the slant TEC ($\TEC$)
 computed from the GNSS carrier phases,
 and its temporal rate of change delta-TEC (dTEC, $\dTEC$). 
The method used to construct the dTEC statistics 
 starts with an ionospheric characterization tool
 written largely in 
 the Python programming language (Python Software Foundation, https://www.python.org/), 
 with supervisory scripts written for the 
 Bash shell program and command language (GNU, https://www.gnu.org/software/bash),
 and the steps followed are as outlined in detail
 in {\APPENDSUP}. 
It analyses a set of RINEX 
 (Receiver Independent Exchange Format, https://igs.org/wg/rinex/)
 files 
 that span either hours or days worth of data,
 and uses the GNSS signal properties listed
 to estimate Total Electron Content (TEC)
 along each signal path through the ionosphere,
 and then also that estimate's variation over time. 
With data from over 300 ground stations available 
 on an on-going basis from CDDIS \cite{CDDIS-NASA},
 and all four GNSS constellations (providing over 130 satellites), 
 a reasonable global coverage can be achieved.
Key steps, methods, and assumptions are these:

\begin{enumerate}[(a)]

\item
 the data source is \cRed{those hourly RINEX files from CDDIS
              that contain data sampled at 30 second intervals},

\item
 the ionosphere is taken to be a thin surface at 350km altitude,
 and the inferred TEC or dTEC values are taken to be
 on the line-of-sight intersection on this surface
 between the ground station and the selected satellite position;

\item
 a Melbourne-Wubbena algorithm is used to flag cycle-slip events
 where signal disruption indicates that TEC (and hence dTEC)
 should be discarded as unreliable.

\item
 subsequently, 
 a normalisation process is applied to align dTEC values obtained 
 from different signal pairs and elevation angles,

\item
 and the dTEC values are binned and histograms are fitted to parameterise 
 the distributions for display and analysis.

\end{enumerate}

%
\subsection{Ionosphere}\label{S-methods-ionosphere}

\cRed{Our assumption of a thin ionosphere
 at a specified altitude
 means that}
 any `line of sight' signal between a satellite
 and a ground station will cross it (intersect it)
 at a well localized intersection point
 (i.e. the ``ionospheric pierce point'', IPP)
 \cite{Klobuchar-1997aes,Sparks-BP-2011rsc}. 
However, 
 since the 350km altitude assumption itself introduces uncertainty, 
 at a later stage we will pixelize the results --
 i.e. assign each result to some fairly coarse angular
 (latitude and longitude) range.
As a result, 
 on the scales considered, 
 the ``thin ionosphere'' assumption has minimal effect.

Since we know the locations of the ground stations,
 and can predict the locations of the satellites based on their navigation data,
 the properties of any signals sent and received
 can be used in an attempt to characterize the behaviour
 of the ionosphere at the intersection point.

For a particular ground station and satellite pair, 
 and for
 the geometry-free combination from two frequency bands,
 we can infer
 the ionospheric slant TEC \cite{Aarons-1982ieee} along the signal path.
In standard TEC units (TECU), 
 and at epoch $t_k$,
 we have that 
~
\begin{eqnarray}
  \TEC(t_k)~=~
  \textrm{TEC}(t_k)
&\simeq
  \frac{1}{40.3}
  \frac{f_1^2 f_2^2}{f_1^2 - f_2^2}
  \left[
    \lambda_1 L_1(t_k) - \lambda_2 L_2(t_k)
  \right]
,
\label{eqn-TECestimate}
\end{eqnarray}
 where $L_1$ and $L_2$ are carrier phases 
 associated with signal frequencies of $f_1$ and $f_2$; 
 and likewise $\lambda_1$ and $\lambda_2$ are their wavelengths.
TEC is measured in $m^{-2}$, 
 with TECU being $10^{16} m^{-2}$; 
 the factor of $40.3$ is 
 $\kappa = c^2 r_e / 2\pi$.
Note that this equation generally neglects any
 residual error term, ambiguities, biases,
 and receiver noise.

We then assume that we can use 
 the change in time of that inferred slant TEC
 as a measure of the variation in ionospheric properties
 (due to ionospheric irregularities)
 at or near where the signal intersected it.
Specifically, 
 the rate of change dTEC
 based on 30 second interval data would be 
~
\begin{eqnarray}
 \dTEC(t_k)
&=
  \left[ 
    \TEC(t_k) - \TEC(t_{k-1})
  \right]
  /
  (30s)
,
\end{eqnarray}
\cRed{leaving us with a $\dTEC(t_k)$ in units of TECU$/s$.}
In what follows we focus primarily on 
 \cRed{what large ensembles of these dTEC values 
 might tell us about 
 the state of the ionosphere,
 whether} globally,
 or as localised in time and space.

%
\subsection{Processing and initial analysis}\label{S-methods-processing}

From an input of GNSS data in the form of RINEX files, 
 the primary output is essentially columns of data specifying TEC or dTEC values,
 their location, time, and station-satellite pair,
 as well as the signal band combination used. 
For a one-day dataset comprised of 30s intervals, 
 as downloaded from CDDIS \cite{CDDIS-NASA,Johnston-RH-2017shignss},
 these can easily encompass several tens of
 millions of data points (typically 55-60 million per day),
 which can then be filtered at will,
 and analysed or visualised as convenient.

For a single day, 
 uncompressed dTEC filesizes are about 6-7GB in size. 
The time taken to generate these dTECs is about one hour on 36-core workstation, 
 although subsequent analysis steps -- 
 which are not so easy to parallelize -- 
 take a similar timescale.
Currently the processing encompasses the GPS, GALILEO, BEIDOU, and GLONASS constellations.
In our analysis that follows, 
 the calculation of dTEC, 
 and its temporal or spatial variation,
 is key to revealing \cRed{GNSS-relevant} ionospheric disturbances.

%
\subsection{Data subsets and slices}\label{S-methods-slices}

The accumulated dTEC data, 
 or specific subsets (``slices'') thereof 
 were then used to create frequency histograms 
 of signed dTEC values.
As indicated in the Preview, 
 although it might seem more straightforward to simply 
 calculate moments of these data samples, 
 in some cases we cannot be sure in advance whether 
 such moments will adequately represent the underlying physics, 
 due to the possible presence of power-law tails 
 with low-value exponents.
Instead, 
 we just
 fit each data slice to a simple empirical model. 
This model was chosen after experimentation with different
 and variously complicated fitting functions.

\begin{figure}
\includegraphics[width=0.9800\columnwidth]{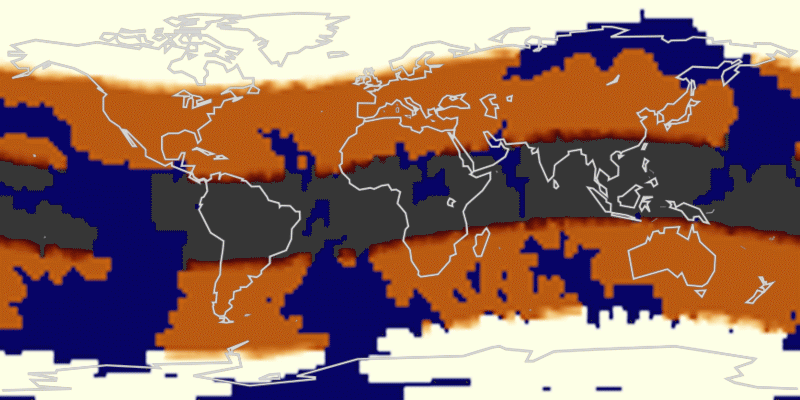}
\caption{Shows a combination of measurement sampling and categorization
 into low, medium, and high magnetic latitudes
 (coloured black, brown, and white respectively), 
 regions where there is no data coverage are dark blue.
The magnetic latitude zones are as determined at the 
 altitude of the ionosphere
 (i.e. not the magnetic latitude at ground level), 
 as based on a sinusoidal approximation to
 the relevant figure(s) shown at the end
 of the WMM2015 \cite{WMM2015} and WMM2020 \cite{WMM2020} reports.
 We use an ionosphere-altitude demarkation of magnetic latitude, 
 because the signal intersection/interaction is in the ionosphere,
 and not at ground-level.
}
\label{fig-mapMagLat}
\end{figure}

For the statistical analysis, 
 the available dTEC data 
 is divided \footnote{Terminology:
   ``dividing''/``division'' refers to a restriction
   along some axis (eg setting T05),
   and a slice is the result of one or several divisions
   (e.g. at T05, lonc:090, mlz:1n, and lpair:*) ...
   We reserve ``bin'' for use when referring
   the assignment of dTEC values into sub-intervals for histograms.}
 into \cRed{subsets (here called ``slices'')}
 along any combination of four separate 
 \cRed{types of division}:

\begin{description}

\item[Hours] each of the 24 hours in a day extends from $0$ minutes $0$ seconds through to 
 $59$ minutes $59.\dot{9}$ seconds\footnote{The GNSS data is supplied in GPS time, 
 which is close to but not the same as UTC (there is an 18s offset 
 as at 2025). However, we only use the data in slices at least an hour long, 
 and our dTECs are based on differences computed at 30s intervals.
This means that for our purposes  
 the offset between GPS and UTC is unimportant,
 and so we choose to label times as being in the widely used UTC, 
 rather than the less well known GPS time.}

\item[Degrees longitudes] 
 we use 30 degree divisions (``dodecants'', or twelfths of the globe),
 and the associated division extends to 15 degrees west of the central value, 
 and up to but not including 15 degrees east of the central value.

\item[Magnetic latitudes]
 separate all latitudes into five divisions,
 low, medium (north and south), high (north and south)
 according to a sinusoidal fit to the magnetic latitude (``mlz''), 
 as depicted -- and justified -- on Fig. \ref{fig-mapMagLat}.

\item[Band pair]
 covers all possible pair-wise combination of the broadcast GNSS
 signals with matched coding; 
 \cRed{as well as some with mixed 
 L1-L2 coding where no L1-L2 matched pair existed, or where no other matched 
 code band pairs were present.}

\end{description}

These divisions, 
 along with their defined labels and values
 are all listed on table \ref{table-divisions},
 which also lists some derived alternate divisions, 
 namely two-hour slices and local (or solar) times.
These allow \cRed{tens of thousands of potential slices --
 for a typical hour,
 where there might be 32 band pairs present, 
 this would be}
 $24 \times 12 \times 5 \times 32$, 
 i.e. over 45 thousand (45k) potential slices.

Owing to the erratic geographic spread of ground stations,
 their signals availability,
 these slices do not always contain enough events
 for consistent sampling and reliable analysis. 
We therefore also compute aggregate properties,
 i.e. based on dividing over only three, 
 two,
 or one of the data divisions.
 This gives us \cRed{many} extra slices to consider. 
Note that -- 
 as explained later --
 we normalise against the different responses of the band pairs, 
 and so most of our analyses will put 
 results from any band pair
 in the same slice; 
 thus the number of slices ordinarily considered is reduced to about 2k.

There is a key tradeoff here,
 which we have tried to bridge by using both fully divided data
 and aggregated data. 
Because of the highly contingent nature of signal availability and reception, 
 and of ionospheric properties,
 at different times or locations,
 we cannot always safely 
 assume that a distribution is perfectly representative, 
 even if the slice contains a large number of dTEC samples.
However,
 it is difficult to sample even the relatively coarse zones 
 adequately if we insist on the quiet conditions needed for a ``nothing happening'' reference.
Notably,
 we find many more recorded events in northern latitude
 (sometimes even eight times more at high north 
  compared to high south); 
 which is a key reason why we choose our divisions 
 to distinguish between northern and southern 
 mid and high magnetic latitudes.
Without such a division, 
 an apparent ``all high latitudes'' result would in fact 
 contain a significant bias towards northern high latitudes.
Likewise,
 aggregated event counts in our longitude divisions
 vary by up to a factor of five between the highest (for lonc:000)
 and the lowest (for lonc:180, 210, 330).

Notwithstanding these sampling concerns, 
 by allowing slices localized in both space and time, 
 we can still track moving disturbances
 \cRed{on a coarse geographical scale}, 
 as well as follow more generic sun-following ionospheric features; 
 thus enabling useful predictive capability.

\setlength{\tabcolsep}{8pt}

\begin{table}
\begin{center}
\begin{tabular}{ || l | c | c || }
\hline
\hline
Category  & Label & allowed values  \\
\hline
\hline
Hours & T    & T00, T01, T02, .., T23 \\
Longitude    & lonc & 000, 030, 060, .., 330 \\
Magnetic && \\
~~~~~~~~latitude & mlz & 0, 1n, 1s, 2n, 2s \\
Band pair      & lpair  & L1CL2C L1CL2W L1CL2X \\
& & L1CL6C L1DL5D L1DL7D  \\
& & L1IL6I L1IL7I L1LL2L  \\
& & L1PL2I L1PL2P L1PL5P  \\
& & L1WL2W L1XL2I L1XL2X  \\
& & L1XL5X L1XL6X L1XL7X  \\
& & L1XL8X L1cL2c L1cL5c  \\
& & L1cL6c L1cL7c L1cL8c  \\
& & L2IL6I L2IL7I L2XL5X  \\
& & L2XL6X L2XL7X L2cL5c  \\
& & L5DL7D L5QL7Q L5QL8Q  \\
& & L5XL6X L5XL7X L5XL8X  \\
& & L5cL6c L5cL7c L5cL8c  \\
& & L6IL7I L6XL7X L6XL8X  \\
& & L6cL7c L6cL8c L7QL8Q  \\
& & L7XL8X L7cL8c \\
\hline
$\S$ UTC 2 hours & D    & D00, D01, D02, .., D11 \\
$\S$ Local time, & & \\
~~~~~~~~2 hours & S    & S00, S01, S02, .., S11 \\
\hline
\hline
\end{tabular}
\end{center}
\caption{Labels used to specify the data divisions. 
Longitude slices
 are centred on the degree position specified, 
 but time slices start at the hour specified.
For magnetic latitudes, 
 we use ``0'' to indicate low latitudes, 
 ``1n'' and ``1s'' for northern and southern mid latitudes, 
 and ``2n'' and ``2s'' for high magnetic latitudes.
\cRed{The band pair list contains some uncoded pairs
 (denoted with a lower case ``c'')
 derived from RINEX v2 files, 
 and a small number of mixed code pairs
 used when no matched pair is available.}
The two 
 lines marked $\S$ (for 2-hour slices), are derived during 
 postprocessing to aid analysis.}
\label{table-divisions}
\end{table}

%
\subsection{dTEC frequency histograms}\label{S-methods-histograms}

Each data slice contains a number of dTEC events (or values  $\dTEC$)
 that match the division criteria used to define it.
These can be binned into frequency histograms
 which usually -- 
 especially at large event numbers --
 are peaked around a centre near $\dTEC=0$
 and have widths less than \cRed{$0.04$ TECU per second} wide.
These frequency histograms provide us with an estimate
 of the probability density function (PDF) $P(\dTEC)$ 
 for the dTEC values in that data slice.

Since the histograms are peaked, 
 and the number of dTEC events is finite,
 they tend to be poorly sampled at larger dTEC values
 and so we restrict our histograms to have bin centres 
 within the range \cRed{$\left( -0.08, 0.08 \right)$ TECU$/s$}, 
 a restriction which still encompasses the great majority of events, 
 even during extremely active conditions.
We split that range into 400 equal width bins, 
 200 bins centred at negative $\dTEC$'s,
 and 200 centred at positive $\dTEC$'s; 
 there is no bin centred at $\dTEC=0$.

In what follows, 
 although we initially create and analyse histograms 
 based on unnormalised (``raw'') dTEC events, 
 we use those initial results
 to motivate and justify 
 a normalisation proceedure for dTEC values.
Once the normalisation factors have been calculated, 
 in our main analysis we then create, 
 fit, 
 and analyse frequency histograms
 based on \emph{normalised} dTEC values,
 and not on the unscaled dTEC.
However, 
 in either case our fitting procedure is unchanged,
 and we describe that next.

%
\section{Fitting and widths}\label{S-fitting}

%
\subsection{Fitting I: the core and the wings}\label{S-fitting-corewings}

\def\GEgoo{\alpha}
\def\GEoff{\alpha_0}

\def\GEgsw{\alpha_1}
\def\GEgsa{\alpha_2}

\def\GEemw{\alpha_3}
\def\GEema{\alpha_4}

\def\GEepw{\alpha_5}
\def\GEepa{\alpha_6}

\def\GEpwo{\alpha_7}
\def\GEpwe{\alpha_8}
\def\GEpwa{\alpha_9}

\begin{figure}
\includegraphics[width=0.90\columnwidth]{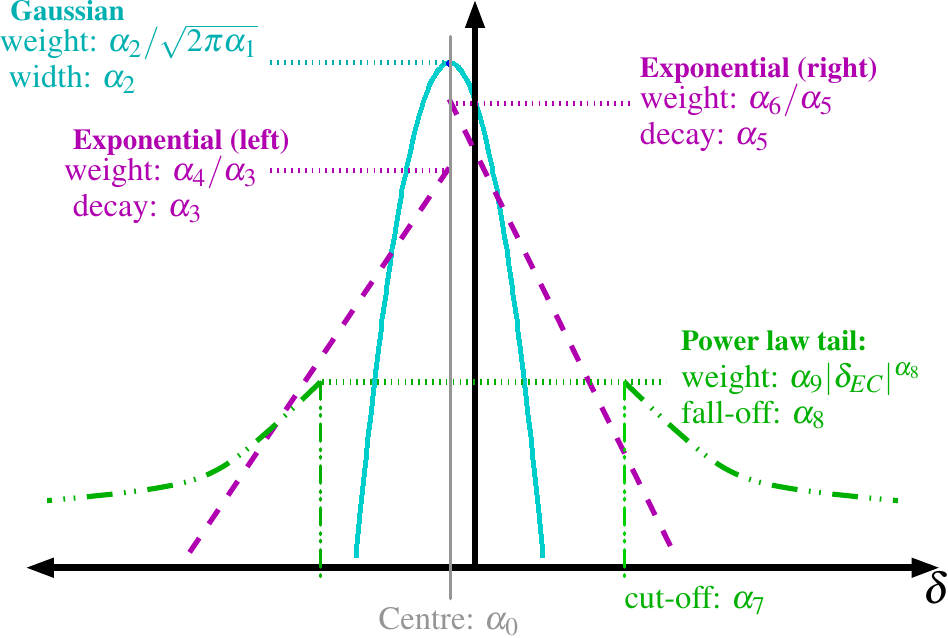}
\caption{A depiction of the four elements of the G2E fit, 
 as well as the power-law tail, 
 using a logarithmic vertical scale,
 and with differences exaggerated for clarity:
(a) the offset of the central peak from the origin at $\dTEC=0$ 
 is given by the parameter $\GEoff$; 
(b) the width and multiplier of the gaussian component (cyan parabola) by 
 $\GEgsw$ and $\GEgsa$; 
(c, d)
 the decay and weighting of the left hand (negative)
 and right hand (positive) exponential components (magenta dashed lines)
 by $\GEemw, \GEema$ and $\GEepw, \GEepa$; and
 (e) the power law components (green dot-dashed lines)
 cutoff, fall-off exponent, and weighting
 by $\GEpwo, \GEpwe, \GEpwa$.
Example histogram fits can be seen in the {\APPENDSUP}.
}
\label{fig-fitscheme}
\end{figure}

The main fitting function contains four parts,
 mainly consisting of a central Gaussian core
 added to two independent exponential decays,
 one in the positive dTEC direction,
 one negative;
 and a shared offset from exact zero dTEC. 
For brevity,
 we will henceforth refer to this as a ``G2E''
 (a Gaussian plus two exponentials) fit, 
 and the various components are depicted 
 on Figure \ref{fig-fitscheme}.
This fit requires 7 real-valued parameters $\{\alpha_i\}$,
 where all but the offset are always positive valued, 
 and where odd indices denote widths, 
 and even indices denote amplitudes.
The combination of functions used in the fit is:
~
\begin{eqnarray}
  p_{G2E}(\left\{ \alpha_i \right\}; \dTEC)
&=
  {\GEgsa}
  G(\GEgsw; \dTEC - \GEoff)
\nonumber
\\
&\qquad
 +
  {\GEema}
  E^{-}(\GEemw; \dTEC - \GEoff)
\nonumber
\\
&\qquad\qquad
 +
  {\GEepa}
  E^{+}(\GEepw; \dTEC - \GEoff)
,
\label{eqn-fitfunction-G2E}
\end{eqnarray}
 where $\dTEC$ is dTEC, 
 $G(\alpha;x)$ is a normalised Gaussian
  with standard deviation $\alpha$ centred at zero, 
 and $E^{\pm}(\alpha;x)$ are one-sided and normalised exponentials
  with width parameters $\alpha_3$ and $\alpha_5$.
One of these ($E^{-}$) is non-zero only for $\dTEC \le \alpha_0$ values, 
 and the other ($E^{+}$) is non-zero only for $\dTEC \ge \alpha_0$ values.
Defining $\delta = \dTEC - \GEoff$, 
 these fitting functions are therefore
~
\begin{eqnarray}
  G(\GEgsw;\delta)
&=
  \frac{1}{\sqrt{2\pi \GEgsw}}
  \exp\left[ - \delta^2 / 2 \GEgsw^2 \right]
  ;
  \qquad \hphantom{\delta=0;}
\\
  E^{-}(\GEemw; \delta)
&=
  \frac{1}{\GEemw}
  \exp \left[ - \left| \delta \right| / \GEemw \right], 
  \qquad
  \delta<0;
\\
  E^{+}(\GEepw; \delta)
&=
  \frac{1}{\GEepw}
  \exp \left[ - \left| \delta \right| / \GEepw \right],
  \qquad
  \delta>0;
\end{eqnarray}
 where $E^{\pm}$ are otherwise zero, 
 except at exactly $\delta=0$ where 
 we replace the value of the exponential function with $1/2$.

When fitting $p_{G2E}$ to a dTEC frequency histogram,
 we set a minimum allowed width for the gaussian and exponential
 components equal to the dTEC resolution scale 
 of \cRed{$0.0002$ TECU$/s$}.

This $p_{G2E}$ is sufficient to accurately characterise
 most distributions with event counts over 10k,
 although note that even on our chosen ``quiet'' days
 (specified later in Section \ref{S-dTECnormalisation}),
 some bins contained event distributions which did not conform
 as well as might be hoped.

Allowing both the centre offset $\GEoff$, 
 a potential smooth central gaussian peak, 
 and the aymmetry of exponentials $E^+$ and $E^-$
 is key to getting an acceptable fit
 on as many sampled distributions as possible; 
 as is the ability to match sharp central peaks
 if the exponential fits dominate.
This is partly because 
 the python routine used (scipy.optimize.curve\_fit) 
 can sometimes --
 although not frequently --
 give poor results for no obvious reason. 
However,
 the large number of data slices requiring fits
 means that 
 they could not all be reasonably inspected and/or patched manually,
 so an automated procedure and a flexible fitting function was needed. 
Here our code first fits the data to both (a) a simple gaussian, and 
 (b) an attempt at that of \eqref{eqn-fitfunction-G2E}; 
 finally choosing the closest match.

When computing the best fits,
 we needed to decide whether or not to assert 
 error bars on the histogram-distributions of sampled dTEC values.
We could, 
 for example, 
 estimate statistical error bars on the basis of the 
 number of dTEC values
 in each of the distribution's (histogram's) internal bins.
However, 
 the naturally much better 
 sampling of the centre would then end up (over-)prioritising a good fit
 to the centre of the distribution 
 at the expense of matching the wings and widths. 
Since here 
 we are more interested in those widths and \emph{wings} of the distributions, 
 we did not specify error bars; 
 thus implicitly asserting equal error bars over the whole range.
It is of course possible that some more systematic proceedure -- 
 or a numerical/ experimental investigation would be preferable, 
 but we leave that for later work.

We characterised the goodness of our fit
 using the summed absolute differences between
 the $P(\dTEC)$ created from the binned data,
 and the fit itself ($p(\left\{ \alpha_i \right\};\dTEC)$), 
 i.e. for bin-centres $\delta_i$, 
~
\begin{eqnarray}
  \textup{Dif}
&= 
  \sum_i \left| P(\delta_i) - p(\left\{ \alpha_i \right\};\delta_i) \right|
.
\end{eqnarray}

See {\APPENDSUP} \ref{S-distribufit}
 for some not-untypical distributions
 as might be seen over a range of event counts.

%
\subsection{Fitting II: the power-law tails}\label{S-fitting-tails}

Although the G2E fitting performs well, 
 it does not capture the behaviour at larger dTEC values, 
 which -- 
 if sufficiently well-sampled -- 
 can be seen to follow a power-law behaviour.
This indicates that these large-dTEC power-law tails are a
 true property of all or some of the data slices, 
 albeit something which is only visible with 
 large enough dTEC counts \footnote{The
  reason for distributions containing
  large dTEC values
  is not clear
 although we believe it is most likely likely due to some physical process. 
However, 
 it remains possible that some or all of the values
 result from technical shortcomings in the RINEX data, 
 or subtle errors introduced by the processing; 
 although it is less clear why such issues might generate such 
 long tailed distributions rather than simpler artifacts.
In any case, 
 here the Tail fits are handled separately, and their contributions
 to the overall widths are optional
 and checked on a case-by-case basis.}.

Consequently, 
 after a G2E fit is done, 
 we also attempt a power-law fit to the \emph{difference}
 between the data and the G2E fit, 
 but only attempting the match in the dTEC range
 i.e. where the discrepancy is large enough
 (more than 75\% of the average of data and fit).
Although these tails do not always appear to 
 have identical fall-offs in the positive and negative $\dTEC$ directions
 (as is also the case for the exponential sub-fits), 
 to minimise the number of additional fitting parameters
 we only fit the sum of the positive and negative tails.

This distribution-tail fitting function is only applied when dTEC 
 values are greater than $\GEpwo$, 
 and with the power-law fall-off parameter $\GEpwe$.
 and amplitude $\GEpwa$,  
 is
~
\begin{eqnarray}
 p_T(\dTEC)
=
 P(\dTEC) - p(\dTEC)
&=
  \GEpwa
  \left| \dTEC \right|^{\GEpwe}
  \*
  H( \left|\dTEC \right| > \GEpwo)
,
\end{eqnarray}
 where $H$ is a Heaviside step function.
This fitting process
 \cRed{which also uses the python routine (scipy.optimize.curve\_fit),}
 does not attempt to match this 
 equation to the usual $P(\dTEC)$, 
 but instead 
 matches a line to $\ln( P(\dTEC))$ 
 as a function of $\ln (\dTEC)$, 
 and error bars on those log-log point positions are 
 assumed to be the same.

The results of such fits on available data  
 indicate 
 that the exponent of the fall-off ${\GEpwe}$ 
 is \emph{not} the same over all data slices, 
 and it can and does vary, 
 typically between values of $-2$ to $-5$.
As noted in the introduction,
 distributions with such power-law tails
 do not always have finite moments,
 the problem being worse for smaller exponents.
This behaviour is why we here use this fitting process, 
 despite its imperfections, 
 rather than simply relying on moments
 such as average, variance, kurtosis, 
 and so on.

A further complication is that such power-law tails 
 do not straightforwardly supply us with either 
 width or amplitude parameters, 
 a point we address later in \ref{S-tailwidths}.
Furthermore, 
 since the tail appears at a very different 
 level of scale and significance, 
 it is not easy to judge the importance  
 of this feature\footnote{Since
   we work with distributions of dTEC values
   automatically generated from ground station date, 
   it is difficult to reliably attribute anomalous features
   to either physical or technical effects; 
   especially given that either -- or both -- 
   can be intermittent, rare, or non-periodic, 
   and either case could also give rise to similar features.
  We do exclude some band pairs (i.e. L1IL6I, L1IL7I) since they 
   seem never to produce usable distributions, 
   and occasionally temporarily remove ground stations from the data 
   if they seem to produce anomalous results (e.g. LAMA in early 2024).
  However, 
   automating such exclusions is challenging, 
   and not unlikely to be fallible, 
   so currently we rely on human observation of the outputs.}:
 is it somehow a key indicator, 
 or an unimportant side effect?

%
\subsection{Ionospheric activity and distribution widths}\label{S-distributionwidths}

A key output we aim for here is a simple aggregate 
 ionospheric activity index
 (or scale),
 which can inform on \cRed{GNSS-relevant} propagation disturbances likely to be experienced
 in a given area and over a given internal of time --
 and therefore on the corresponding impact.

Since dTEC variances \cite{Forte-2005jastp}
 can be correlated with scintillation measures, 
 and our fitted distributions also have variances
 (and standard deviations), 
 we ensured that the fitted parameters for the widths
 are designed to ensure they correspond \emph{directly}
 to the standard deviation of the distribution
 (whether Gaussian or exponential).
\cRed{Then we take a weighted average of the gaussian
 and exponential width contributions,
 using a quadratic combination of the two exponentials
 that allows for both similar and dissimilar fit parameters:
~
\begin{eqnarray}
  W_{2E}
&= 
  \alpha_A
 -
  \frac{1}{2} \alpha_E
  \left[
    \alpha_A
   -
    \frac{\alpha_r}{2}
    \left(
      \alpha_D + \alpha_W . \alpha_r
    \right)
  \right]
  W_{dT}
\\
&=
  2
  \frac{\GEgsw \GEgsa}      
       {\GEgsa + \alpha_A}  
 +
  \frac{ W_{2E} }
       {\GEgsa + \alpha_A}  
,
\end{eqnarray}
with 
 $\alpha_A = \GEema + \GEepa$, 
 $\alpha_D = \GEemw - \GEepw$, 
 and 
 $\alpha_r = \left(\GEepa - \GEema\right)/\left(\GEema + \GEepa\right)$.}
This weighted average is chosen so that the aggregate width is 
 (or can be) 
 dominated by the most significant component(s) of the fit function, 
 and is not, 
 for example, 
 unduly affected 
 by a very wide but very-low weight Gaussian component.

%
\subsection{Tails and widths}\label{S-tailwidths}

The ``distribution width'' measure 
 $\WdT$ introduced above
 does not include any width-like contributions from 
 the fits to the power-law tails.
This is because 
 the tails have neither  
 a natural width scale, 
 nor any natural amplitude scale.
Further, 
 they in no way (ever) contribute a significant fraction 
 of the number of dTEC values sampled.
However, 
 that does not mean that the tail fitting parameters do not contain 
 useful information, 
 but only that it is not straightforward to determine to what extent
 they should (or might) contribute to a combined width.

Since $\WdT$
 is intended as a proxy for ionospheric activity, 
 the ability to consider adding an extra pseudo-width component 
 calculated from the tail fits 
 (from e.g. \eqref{eqn-tailfit-pwidth} and \eqref{eqn-tailfit-amplitude})
 seems desirable.
This could capture and incorporate information on the varying magnitude of dTEC
 under changing ionospheric conditions,
 thus potentially increasing the accuracy and utility
 of $\WdT$ as an ionospheric scale
 (from which the likelihood of ionospheric propagation disturbances
 in a given region and over a given temporal interval can be deduced).

We now construct width-generating tail parameters
 by comparison to typical values of fit parameters 
 as seen on our ``quiet days'' data, 
 and after our dTEC normalisation procedure.
Although both of these considerations
 are defined in following sections
 we deliberately address the construction here so as to keep all 
 of the fitting and width calculations together
 in this section.
Notably,
 we assume
 that the tail behaviour is of roughly equal importance 
 as any of the gaussian or exponential G2E contributions; 
 for this we need to compute a width-like parameter
 and an amplitude-like one that vary over a similar range
 to the G2E ones.

\def\wTail{\omega_T}

First,
 we start from the observation that 
 exponents $\GEpwe$ typically vary between $-2$ (broad) and $-5$ (narrow), 
 whilst typical gaussian and exponential widths on 
 our ``quiet days''
 usually have a minimum of about $0.05$.
Thus 
 here we use an ad hoc procedure where 
 we assign an effective ``tail width''
 parameter to be 
~
\begin{eqnarray}
  {\wTail}
&=
  \cRed{\frac{1}{100 \left| \GEpwe \right|}}
,
\label{eqn-tailfit-pwidth}
\end{eqnarray}
 where \cRed{dividing by 100}
 ensures that ${\wTail}$, 
 like the other widths, 
 also ranges from \cRed{$0.002$ TECU$/s$} upwards.

Second,
 to construct an effective amplitude-like tail parameter (or weighting) 
 we integrate the power-law tail from the cut-off (at $\GEpwo$)
 to infinity, 
 and compare that to a tail normalised to have unit area.
\cRed{To normalise systematically, 
 we set a tail cutoff scale of $p_T=0.01$, a value
 which is similar to typical fitted cut-offs, 
 and also brings the result 
 into same range or values as that for the gaussian and exponential amplitudes.}
Recalling that $\alpha_9$ is the amplitude of the tail feature, 
 this rescaled normalisation approach
 gives us an effective ``tail amplitude'' of
~
\begin{eqnarray}
  \beta_T
&= 
  \cRed{\GEpwa
  \frac{ \left| \GEpwo p_T \right|^{\GEpwe+1} }
       { \GEpwe+1 }}
.
\label{eqn-tailfit-amplitude}
\end{eqnarray}

As a result, 
 should we wish to apply it, 
 we would add the product $\beta_T {\wTail}$ 
 of these two estimates
 to $\WdT$
 to obtain a width-like quantity
 that took these power law tails into account
 with a broadly similar emphasis to the true width estimates:
~
\begin{eqnarray}
  {\WdTs}
&=
  \WdT + \beta_T \wTail
.
\end{eqnarray}

Of course, 
 these initial estimates as to how to incorporate
 the significance of the power law tails are tentative, 
 and in future work we intend to place them on
 a firmer footing.

%
\section{dTEC Harmonisation/ normalisation}\label{S-dTECnormalisation}

Recall that
 the code calculates TEC values from all possible band-pair combinations
 received by any one ground station from any one satellite.
These uncalibrated TEC estimates give rise to dTEC values that 
 contain two significant sources of systematic bias.
These biases need to be considered, 
 and compensated for, 
 and once this is done 
 we can aggregate the compensated dTEC values 
 into larger slices.

In order to achieve this harmonisation
 of the individual dTEC values based on different parameters, 
 ideally we would like to use as a reference 
 a well behaved set of dTEC data with a minimum of 
 extreme or unusual features.
In this way we would hope that it will be dominated by
 correctable biases, 
 rather than being distorted by (e.g.)
 artifacts caused by a poor sampling of unlikely phenomena.
Further,
 our primary interest here is in recognising, 
 predicting, 
 or categorizing 
 interesting, unexpected, or disruptive events.
That is, 
 we want to be able to flag up events or behaviours that are somehow
 \emph{different} to a benign and quiet ionospheric background.
We therefore address this issue 
 before proceeding to our method for normalising dTEC values.

%
\subsection{Reference choice: ``quiet days''}\label{S-quietreference}

The ionosphere is rarely ``quiet'' in any meaningful sense. 
For example, 
 if we use the Potsdam $\KPI$ index as an indicator for quiet conditions,
 note that on only three days between 2020--2022 was it zero, 
 and on only three more was it its next-lowest value $0.3333$.
To this purpose,
 we \cRed{primarily consider the} 
 days 006 and 007 in 2022 as our benchmark for reasonably quiet ionospheric conditions.
Furthermore,
 the choice of quiet days 
 away from equinoctial conditions,
 is a way to increase sensitivity towards enhanced ionospheric activity
 at equatorial latitudes during equinoctial conditions.
\cRed{However,
 in order to enable consistent normalisation across long periods, 
 where older data might well be 
 contained in RINEX v2 files, 
 we also use comparably quiet two-day periods,
 namely  2015/328 and 329, 
 and 2020/314 and 315. 
These older reference periods,
 both in the month of November, 
 enable us to also normalise the uncoded or mixed-coded bandpairs
 that older data files force us to consider.
Note that we normalise the three periods individually, 
 then chose the normalisation result for a bandpair from the period
 with the higher number of dTEC samples.
With these benchmarks,}
 we can then rate any other day or days -- 
 or other data slices from other days with respect to this reference; 
 all without 
 having to invent, guess, or intuit 
 some idealised or theoretical reference situation.

Nevertheless, 
 even though these quietest $\KPI$ days (with $\KPI = 0$)
 do have significantly less dTEC activity 
 than most, 
 there is still some,
 and there are still spatially or temporally localised features that 
 can appear in the data.
Thus as long as we are not interested in subtle distinctions 
 (e.g. between ``quiet'' and ``almost quiet''),
 and instead focus on more active ionospheric behaviour,
 our reference choice should be seen 
 primarily as a \emph{pragmatic} one, 
 rather than an attempt at perfection.
In principle we might even dispense with the $\KPI$ index as a
 proxy for quietness,
 and instead trawl all available data for relevant slices 
 to find the minimal variation (in dTEC).
However,
 such a project is currently beyond our computational resources.

In the next subsection we describe how we use the 
 different fitted widths for different band pairs 
 during this quiet reference period to normalise out 
 that difference, 
 enabling us to treat all band pairs as effectively equivalent.

%
\subsection{Normalisation of dTEC values}\label{S-normalisation}

\def\losPath{L_p}

The first potential bias is
 the geometry of the LOS between satellite and ground station:
 notably, 
 LOS paths at a lower angle traverse the ionosphere --
 and any irregularities in it --
 at an oblique angle, 
 and thus spend a longer time (or distance ${\losPath}$)
 under its influence \footnote{We assumed earlier that our 
  ionosphere could be approximated as a thin shell 
  for the purposes of localizing an intersection on a coarse scale; 
  but of course the ionosphere is not thin, 
  and any signal will interact with it over some finite distance.
  Hence,
  this problem needs to be handled in order to be able to
  compare different dTEC values 
  obtained from ray paths intersecting
  field-aligned irregularities in different ways.}.
This gives such signals more opportunity to accumulate time-dependent perturbations, 
 thus potentially affecting the estimated dTEC.
In addition,
 the TEC estimates used to calculate the dTEC itself
 are based on IPPs which are further apart.

However, 
 it is not straighforward to justify a specific compensation or correction method
 on physical grounds.
For example, 
 if the dTEC was largely due to one brief local event, 
 then we should not expect to apply a correction; 
 whereas if the final dTEC resulted from a diffusion-like process
 then the correction should be proportional to $\sqrt{\losPath}$;
 but if drift-like then the correction should be proportional
 to just ${\losPath}$.
Worse, 
 different receive events may produce dTEC values which 
 have a diversity of histories, 
 each containing different contributions from a variety of perturbation mechanisms. 
Accordingly in what follows we take a data-driven, 
 distributional approach to ensure that the statistics of dTEC events
 at different path-angles match as closely as possible.

The second potential bias is that
 dTEC values computed on the basis of different band pairs
 may have different behaviour; 
 and indeed if we fit data slices 
 that only differ by their specification to different band pairs, 
 we see fitted widths 
 \cRed{which might even vary by a factor of two or more}.
This has a relatively straightforward fix --
 we can do band pair specific fits to our reference ``quiet days''
 as discussed above, 
 then use those to compute scalings for the raw dTEC values  
 which we can apply before slicing and fitting the data.

We now specify the two processes
 that enable us to treat the resulting corrected dTEC values
 are independent of both path-angle (or length) and band pair combination.

%
\subsubsection{Path-angle correction}\label{S-norm-pathangle}

Here we take a large set of data
 (i.e. a pair of quiet days),
 and create a set of dTEC frequency histograms
 distinguished only on the basis of path-angle, 
 \cRed{namely by using
 the cosine of the intersection angle of the LOS path with the ionosphere
 (``path-cosine'', $C$).
The distinct input set for each such ``cosine-specific'' histogram
 were determined after rounding the path-cosines to 
 two decimal places.
Then,
 for the dTEC bins in each cosine-specific histogram,
 we discretised the logarithm of dTEC values (in TECU$/s$) to two decimal places.}

\def\picos{\eta}

\begin{figure}
\includegraphics[width=0.753200\columnwidth]{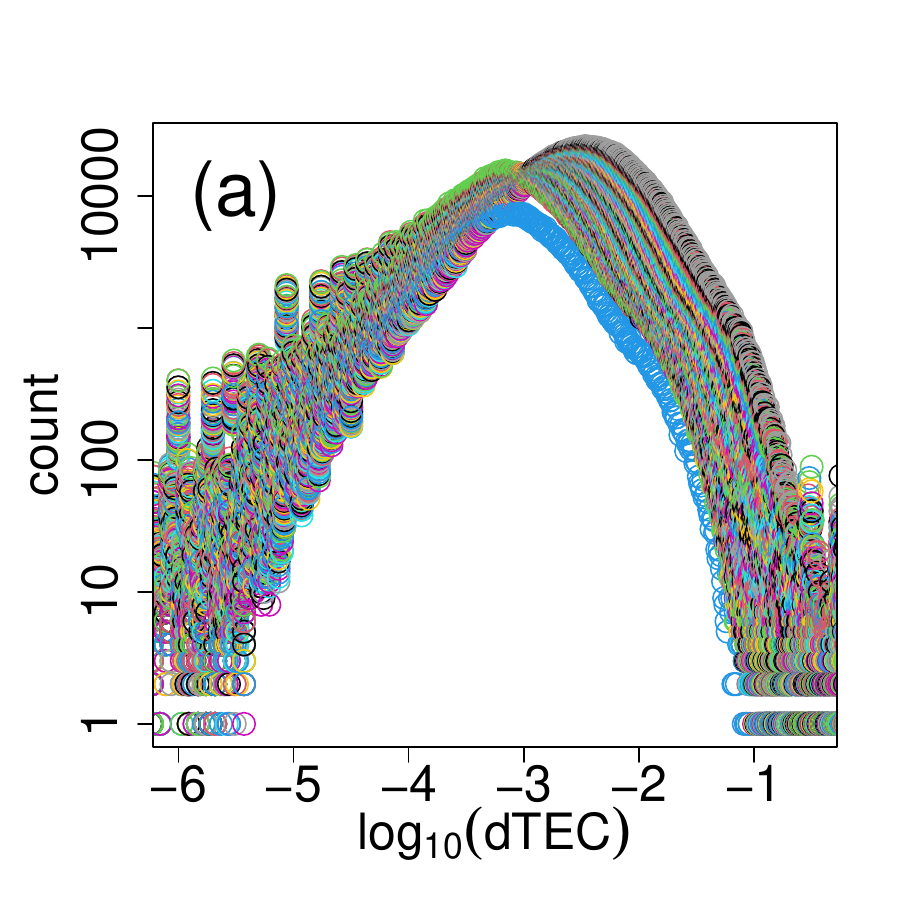}
\includegraphics[width=0.753200\columnwidth]{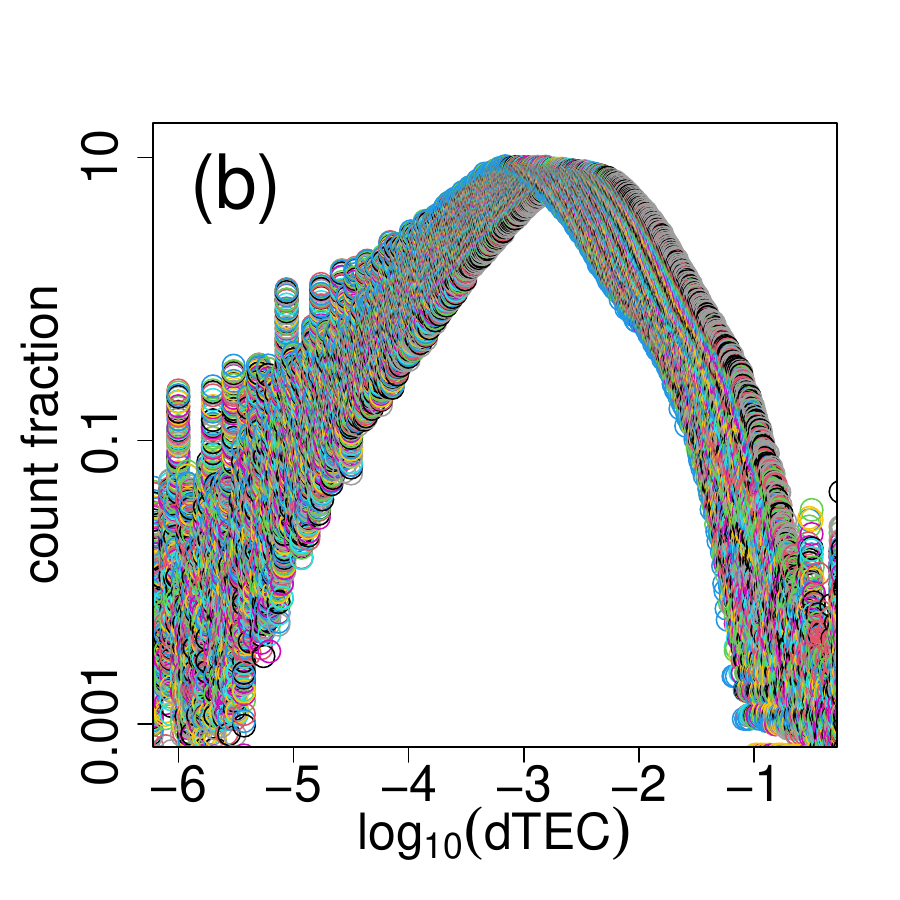}
\includegraphics[width=0.753200\columnwidth]{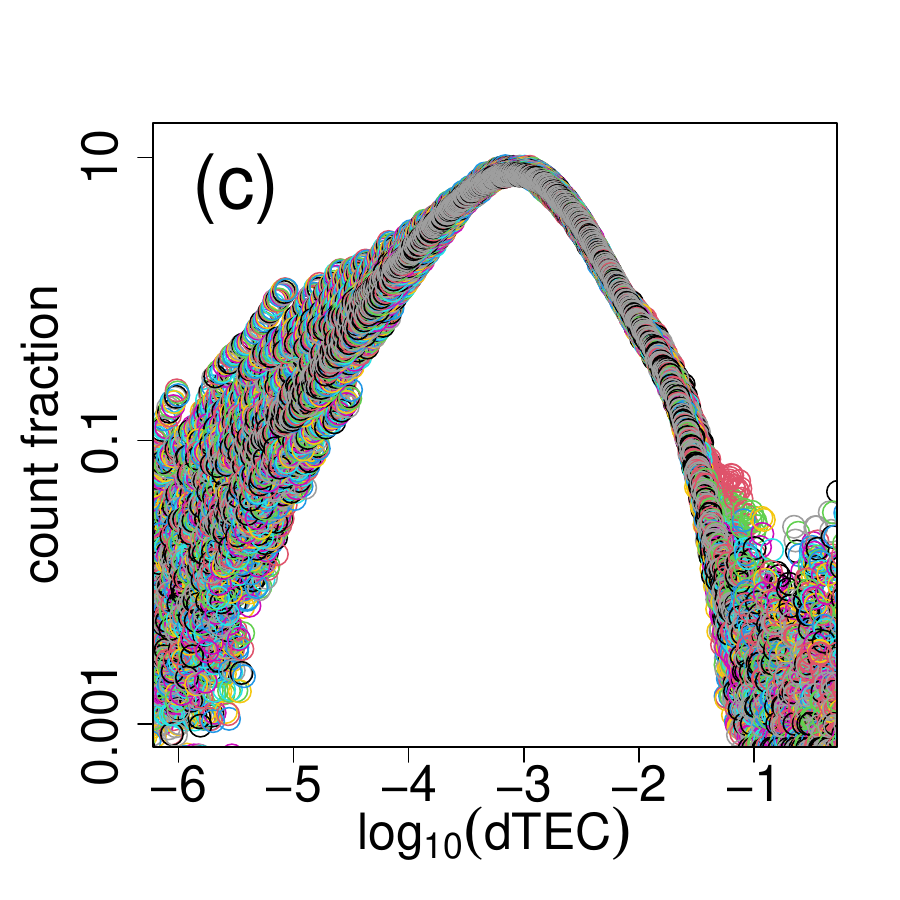}
\caption{Histogram-based plots with different path-angles as different colours, 
 plotted using coloured markers and logarithmic scales to improve visibility.
On the left we see data for entirely uncorrected dTEC values
 and unnormalised histograms, 
 in the centre we see it with normalised histogram data, 
 and on the right with dTEC values additionally scaled by $C^{1.60}$.
Over the most likely ranges of log(dTEC), 
 i.e. roughly between \cRed{$0.004$ and $0.012$ TECU$/s$},
 we see that the 
 histograms for different $C$ are now rather well matched.
This data comprises all that from our quiet 006/007 days in 2022, 
 but the same behaviour was also seen for our notional ``typical'' 
 day 319 in 2022, as well as the active days 113/114 in 2023.
}
\label{fig-reference-icoscorrection}
\end{figure}

If we overlay the resulting set of histograms 
 we see that they are similar in form, 
 but not identical -- 
 notably that they get wider as the path-cosine becomes smaller
 (see Fig. \ref{fig-reference-icoscorrection}).
We found that multiplying the dTEC values at a given path-cosine
 by a power of that cosine can improve the similarity between 
 \cRed{the different cosine-specific histograms.}
\cRed{By minimising mean-squared differences
 between the central regions of the distributions,
 the best match was found to be at a power of $\eta = 1.60$,}
 although the matching was not highly sensitive to the exact value.
This means we will adjust the raw dTEC values to corrected ones,
 in analogy with mapping functions, 
 as
~
\begin{eqnarray}
  {\dTEC}_\textup{(corrected)}
&=
  {\dTEC}_\textup{(raw)}
  \,
  {C}^{\picos}
.
\end{eqnarray}
The steps in the process to justify this correction mechanism are shown
 on Fig. \ref{fig-reference-icoscorrection},
 where
 we see that we can 
 remove the bulk of path-angle effects for the important -- 
 and most frequent -- 
 range of binned dTEC values, 
 i.e. those in the range \cRed{$0.004$ to $0.012$ TECU$/s$}.

It is interesting that the best match power ${\picos}=1.60$ 
 does not fall into any of the no-correction (where we would have ${\picos}=0$),
 diffusion (${\picos}=\tfrac{1}{2}$), 
 or drift (${\picos}=1$) cases we suggested above.
Indeed, 
 it would seem that the ${\picos}=1.60$ indicates that
 however irregularities in the ionosphere  
 affect the eventual dTEC, 
 the effect is enhanced at lower path-angles.
This could be interpreted to mean that dTEC disturbances are 
 more pronounced if passing through ionospheric irregulatities
 more horizontally, 
 but reduced if the path is closer to vertical.
Remember, 
 however, 
 that our ${\picos}=1.60$ results from a statistical comparison, 
 and so can (or will) encompass the effects of 
 many possible perturbation histories.

%
\subsubsection{Band pair scaling}\label{S-norm-lpair}

\def\BPC{B}

The band pair used in any dTEC computation is an important consideration,
 because the TEC (and hence dTEC) values are uncalibrated, 
 and are computed based on band pairs 
 that have frequency differences that
 span 
 an order of magnitude --
 $\Delta f$ for L1CL2C is 356MHz, 
 but for e.g. L7QL8Q it is only 15.345MHz.
Further,
 the frequency differences fall broadly into two groups, 
 those with `large $\Delta f$ (i.e. $> 250$MHz), 
 and those with small $\Delta f $ (i.e. $< 120$MHz).
This grouping 
 is simply an artifact of the transmission frequencies used
 by GNSS satellite constellations considered here.

To compute the scaling factor for a given bandpair we
 start with the complete set of raw dTEC values from our reference data.
We then
 (a) select only those with the correct bandpair, 
 (b) apply the desired path-angle correction, 
 and
 (c) calculate fits and hence the $\WdT$ for the bandpair,
 and finally
 (d) use the ratio of this bandpair-specific $\WdT$ with
 \cRed{that from the standard L1CL2C bandpair},
  to get the scaling factor ${\BPC}$.
\cRed{We can see some of these normalisation factors
 on Table \ref{table-lpnorms},
 and the full table is given in {\APPENDSUP} A.9.}

Thus,
 with these bandpair scalings calculated, 
 the normalised dTEC values we need 
 combine the path-angle correction and the scalings, 
 so that 
~
\begin{eqnarray}
  {\dTEC}_\textup{(normalised)}
&= 
  {\dTEC}_\textup{(corrected)}
  \, .
  {\BPC}({\picos})
\\
&=
  {\dTEC}_\textup{(raw)}
  \, .
  {C}^{\picos}
  \, .
  {\BPC}({\picos})
.
\end{eqnarray}
We can now apply this normalisation procedure 
 to each new set of dTEC values, 
 aggregating them into histograms that can be fitted
 as described in section \ref{S-fitting}.
\cRed{As a result we can take any set or subset 
 of raw dTEC values we have to hand,
 and normalise them 
 so that the values
 become elevation and band pair agnostic, 
 enabling better comparisons to be made.}

\setlength{\tabcolsep}{8pt}

\begin{table}
\begin{center}
\begin{tabular}{ || l | c || }
\hline
\hline
Band Pair  & ${\BPC}(1.60)$  \\
\hline
\hline
L1CL2C & 1.0000 \\
L1CL6C & 1.1482 \\
L1PL2P & 1.0174 \\
L1PL5P & 1.2627 \\
L1WL2W & 1.2596 \\
L1XL5X & 1.1766 \\
L1XL7X & 1.2355 \\
L1XL8X & 1.2732 \\
L2IL6I & 1.1848 \\
L2IL7I & 1.4405 \\
L2XL5X & 0.4862 \\
L5QL7Q & 0.4479 \\
L5QL8Q & 0.3839 \\
L5XL7X & 0.3607 \\
L5XL8X & 0.2823 \\
L6IL7I & 0.5645 \\
L7QL8Q & 0.4193 \\
L7XL8X & 0.2844 \\
\hline
\hline
\end{tabular}
\end{center}
\caption{\cRed{A sample of the computed dTEC normalisation scalings for the 
 most prevalent band-pairs. 
 as based on aggregated data from the ${\KPI}$ ``quiet'' days 006 and 007 of 2022,
 as well as 215/328 and 329, 
 and 2020/314 and 315.
A full table is given
 in {\APPENDSUP} A.9.   
Here ${\BPC}(1.60)$ is the band pair scaling 
 needed to align the fitted widths when the 
 path-angles correction was applied.}
}
\label{table-lpnorms}
\end{table}

%
\section{UTC versus local time}\label{S-g2e-utcvslocal}

Before proceeding to the calculation of our index $\LdT$,
 it is instructive to 
 analyse multi-day sets of data 
 and see how the ionosphere responds to 
 the day/night cycle.
To do this we 
 consider the two distinct temporal bases
 on which we might divide the dTEC measurements:
 either by universal time (UT) or local time (LT).
We now proceed to compare representations of the
 data looked at in both ways; 
 but in order
 to align with our $30^\circ$ longitude divisions (dodecants), 
 we base these comparisons on slices containing \emph{two} hours of data, 
 rather than our typical minimum -- a single hour of data.

\begin{figure}
\includegraphics[width=0.99\columnwidth]{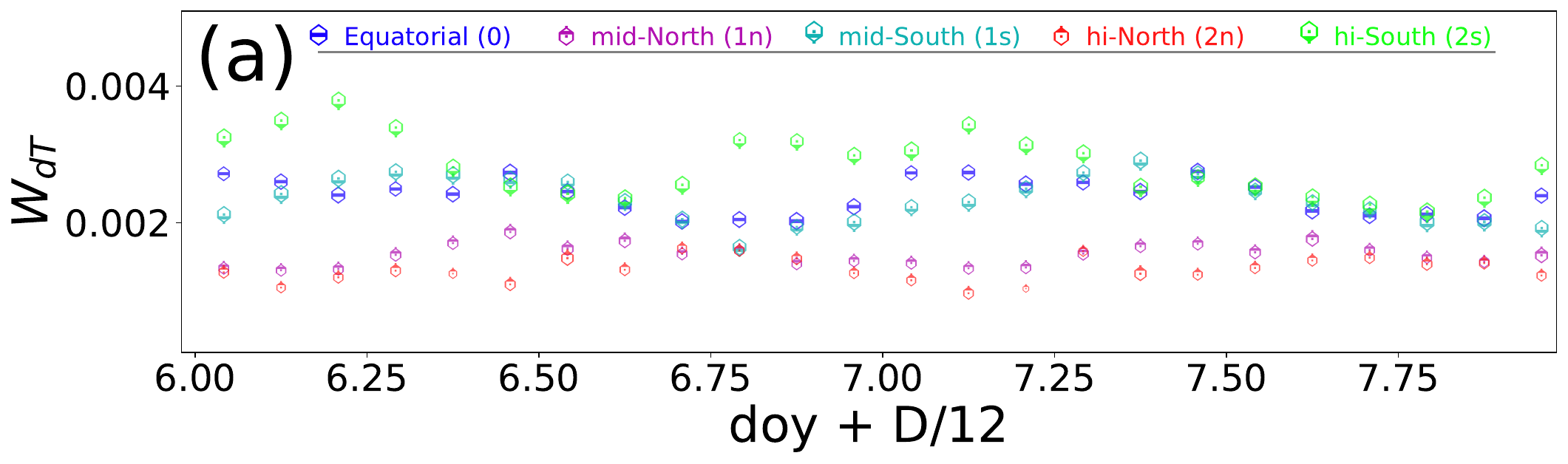}
\includegraphics[width=0.99\columnwidth]{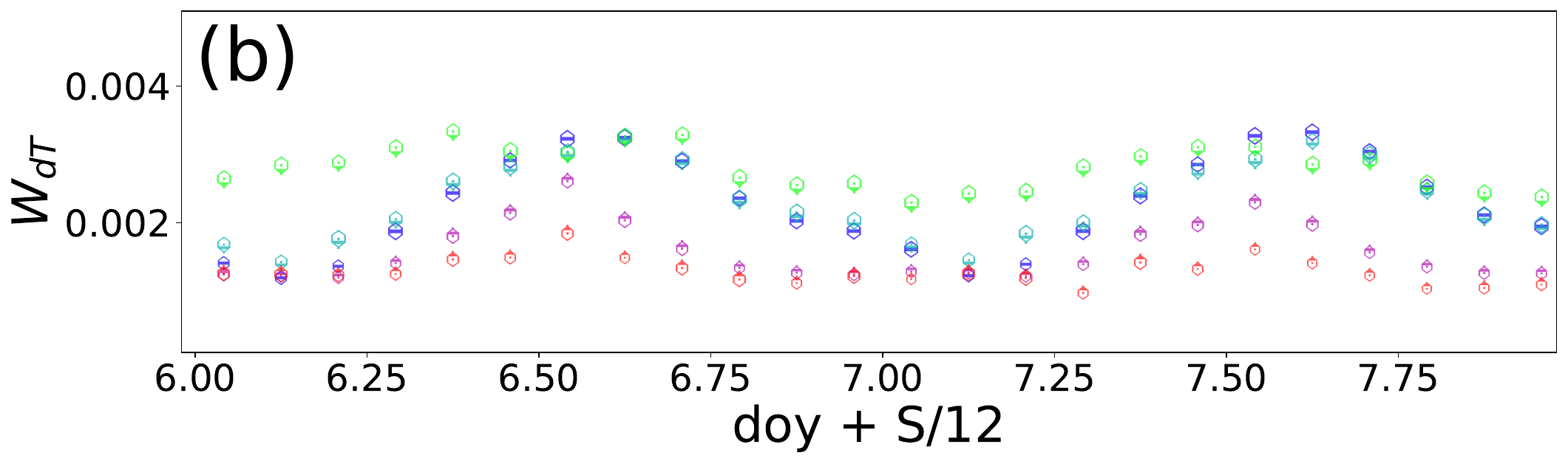}
\caption{The globally-aggregated 
 weighted widths $\WdT$
 shown hour-by-hour for the two quiet days in 2022, i.e.  006, 007.
In (a) we bin the dTEC events in two-hour slices ``D'' by UTC, 
 but in (b) we show them by local/solar time two-hour slices ``S''.
Widths are indicated for each magnetic latitude zone individually,
 and smaller markers indicate widths based on poorer fits to the data.
}
\label{fig-WdT2022-006-007}
\end{figure}

\begin{figure}
\includegraphics[width=0.99\columnwidth]{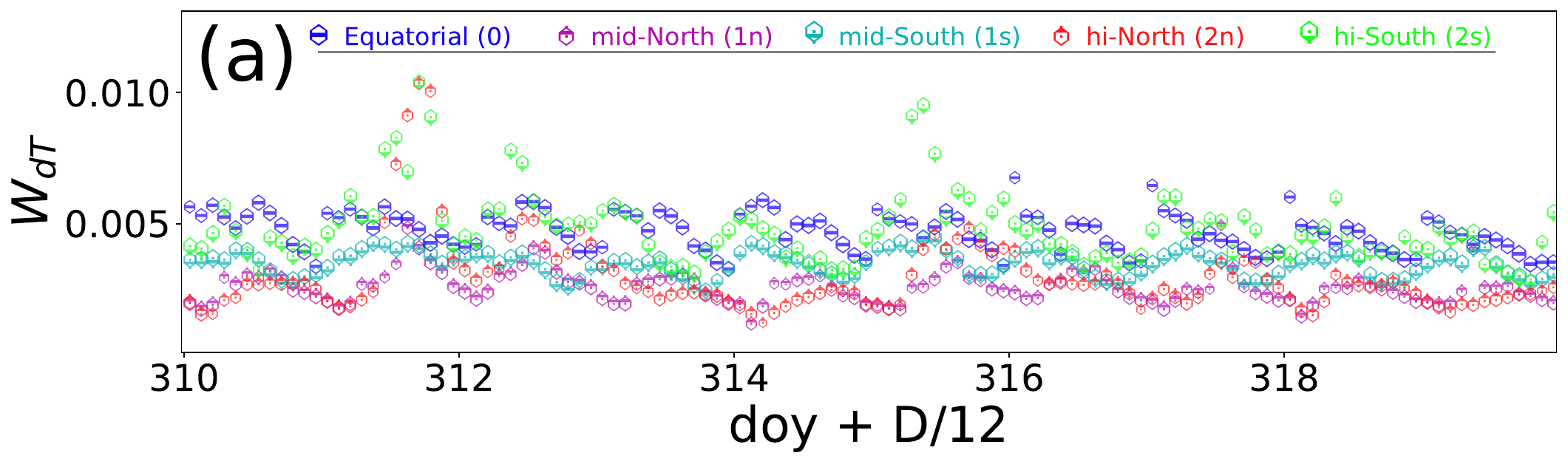}
\includegraphics[width=0.99\columnwidth]{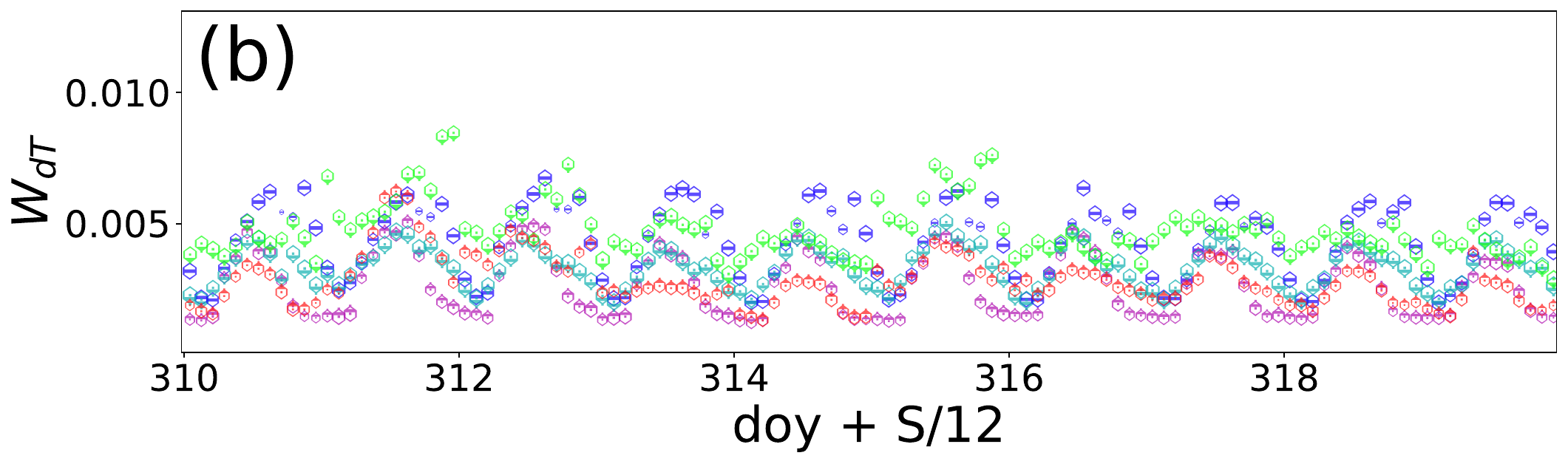}
\caption{The globally-aggregated 
 weighted widths $\WdT$
 shown hour-by-hour for a ten-day period in November 2022.
In (a) we bin the dTEC events in two-hour slices by UTC, 
 but in (b) we show them by local/solar time.
Although these plots contain a great wealth of detail, 
 here we only intend them to indicate general trends
 and behaviours.
The take-home message here is simply the increased regularity and periodicity 
 of the local time data slices, 
 especially at low latitudes.
Widths are indicated for each magnetic latitude zone individually
 and smaller symbols indicate poorer fits.
}
\label{fig-WdT2022-31x}
\end{figure}

\begin{figure}
\includegraphics[width=0.99\columnwidth]{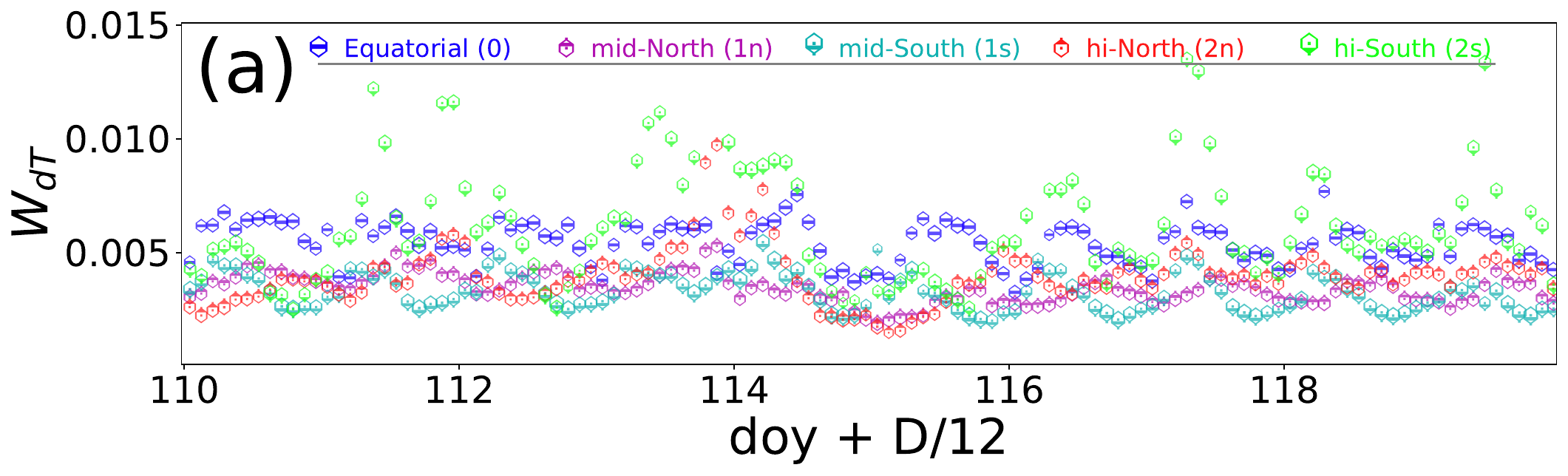}
\includegraphics[width=0.99\columnwidth]{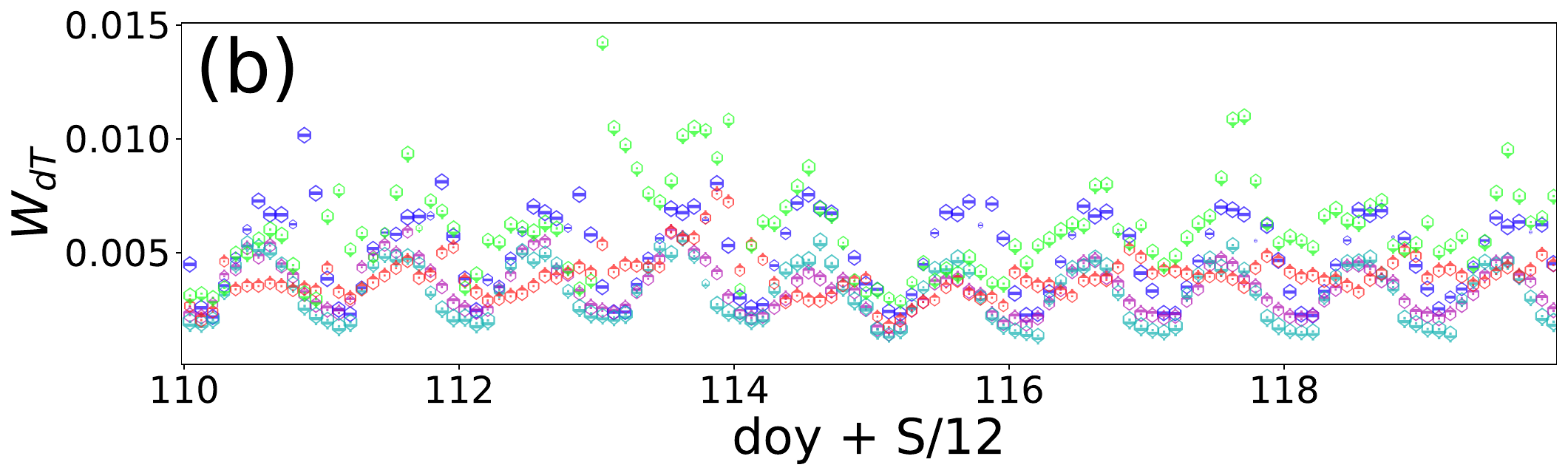}
\caption{The globally-aggregated 
 weighted widths
 shown hour-by-hour for an active ten-day period in April 2023.
In (a) we bin the dTEC events in two-hour slices by UTC, 
 but in (b) we show them by local time.
Although these plots contain a great wealth of detail, 
 the take-home message here is simply the increased regularity and periodicity 
 of the local time data slices, 
Widths are indicated for each magnetic latitude zone individually,
 and smaller symbols indicate poorer fits.
 especially at low latitudes.
}
\label{fig-WdT2023-11x}
\end{figure}

In Figs. \ref{fig-WdT2022-006-007}, \ref{fig-WdT2022-31x}, \ref{fig-WdT2023-11x}
 we compare three different sequences of day-data 
 organised by either universal time (UT) or local time (LT).
In all three sequences, 
 we see that by LT
 the temporal variation 
 has a more regular and more pronounced modulation
 as compared to that aggregated by UT; 
 whilst noting that the LT version
 also moderates the effect of the uneven geographic sampling.

Notably, 
 we see here that the LT classification 
 is likely to be invaluable when (e.g.)
 making predictions based on day/night cycles.
However, 
 the recasting into LT does obscure 
 temporally localised events
 (e.g. in 2022 doy 311 on Fig. \ref{fig-WdT2022-31x},
  and 2023 doys 113/114 on Fig. \ref{fig-WdT2023-11x} ) are obscured.

This use of local time aggregation is most useful when 
 looking at a global picture, 
 but it is also possible to break down that coverage 
 into magnetic latitude and longitudes, 
 as we do in the next section.
However, 
 it is also possible to disassemble
 the distribution widths themselves
 into their gaussian, exponential, or power-law subcomponents.
Such detailed breakdowns will be discussed elsewhere,
 but in the following we will concentrate on 
 latitude and longitude divisions, 
 and in a log scale index based on the fitted widths
 that will aid end-user interpretation.

%
\section{Log-scale Index: $\LdT$}\label{S-L-index}

Although it is scientifically interesting to look at 
 a very detailed breakdown of the dTEC statistics, 
 what a forecaster or \emph{end-user} will want to know 
 is arguably restricted to two things:
 how bad is it (was it) in general, 
 and what was it like at worst?
The distinction is relevant since 
 GNSS-effective activity tends to be localised, 
 all-longitude averages typically contain large regions
 of low activity even on active days, 
 thus meaning an average measure does not represent 
 the actual strength of the disruption where it was 
 strongest.
Accordingly,
 although an average activity measure is nevertheless useful, 
 we also would like to know the maximum, 
 albeit being under the caution that relying on a single fitted width 
 might sometimes be misleading 
 and that some longitudes have better coverage than others.
Further, 
 since we are now interested in temporally localised behaviour,
 here we show results based data arranged by UTC and not by local time, 
 although this is not a requirement for computing an $\LdT$ measure.

Although the combined width measures $\WdT$ or ${\WdT}^*$
 are scientifically useful, 
 being in units of dTEC/s, 
 for more general usage it is likely that 
 a logarithmic scale, 
 could provide a better summary for any wider audience.
This is because logarithmic scales can encompass a wider range of activity
 without a similarly wide range of values, 
 where the wide range provides detail unnecessary 
 and distracting for the non-specialist.
Further, 
 careful choice of scaling can reduce our new 
 logarithmic ``$\LdT$'' activity measure (defined below)
 to easily-quotable integer values, 
 removing the need 
 to quote activity using decimals, 
 such as we would have to do if quoting $\WdT$ values.

As noted above, 
 choosing appropriate scalings will help make the 
 $\LdT$ values more user-friendly.
Here we pick a scale intended to have a minumum value of 0 --
 except, possibly, under extremely quiet conditions --
 and which increments by 2 for each doubling in width.
Note, 
 however, 
 that here any ``user experienced effects''
 most directly related to our dTEC index 
 will be positioning error and scintillation strength; 
 but this index has not yet been formally calibrated against those.

We saw previously that 
 the G2E fitting process returns three width parameters, 
 each with an amplitude weight, 
 being for the
 one (double-sided) gaussian, and two one-sided exponentials.
We then evaluate the combined weight using $\WdT$
 and compute our logarithmic index as follows:
~
\begin{eqnarray}
  \LdT
&= 
  2 ~ \log_2 \left(500 \, \WdT \right)
.
\end{eqnarray}

Here the factor of \cRed{500 inside the log sets our baseline width
 of a quiet-days L1CL2C bandpair}
 to give an index of zero,
 and the multiplier 2 means that increments of 2 
 imply a doubling of $ \WdT$.
The range spanned by the $\LdT$ index
 depends on the slice of data for which $\LdT$ is being evaluated.
For a one hour slice based on one magnetic latitude zone and all-longitudes, 
 the typical variation is between 2 for ordinary quiet conditions, 
 and 6 for with significant activity; 
 however for smaller slices
 (e.g. restricting to a single 30 degree longitude dodecant)
 the variation is greater -- 
 the lower value can drop below zero, 
 and the highest value might exceed 8.

To indicate both the typical level of activity 
 as well as report on its extremes, 
 we divide the dodecant set of longitude slices 
 (and their computed $\LdT$ values) 
 up into three parts
We:

\begin{enumerate}[(a)]

\item
 estimate a maximum by averaging 
 the two largest $\LdT$ values in the set,

\item
 estimate a minimum by averaging 
 the two smallest $\LdT$ values in the set,

\item
 estimate a midpoint by averaging 
 the remainder of the $\LdT$ values, 
 i.e. the middle eight values.

\end{enumerate}

The result is that we can report 
 the presence of large-scale irregularities
 forming in the ionosphere
 in some latitude band
 in terms of both a typical value
 (the average of the non-extreme values)
 and its likely maximum excursions.
Note that if all dodecants fail to return a valid $\LdT$, 
 we simply use fewer values for the midpoint average; 
 this eventuality is most common in southern polar latitudes 
 due to the sparse distribution of ground stations.

\begin{figure}[ht]
\includegraphics[width=0.99\columnwidth]{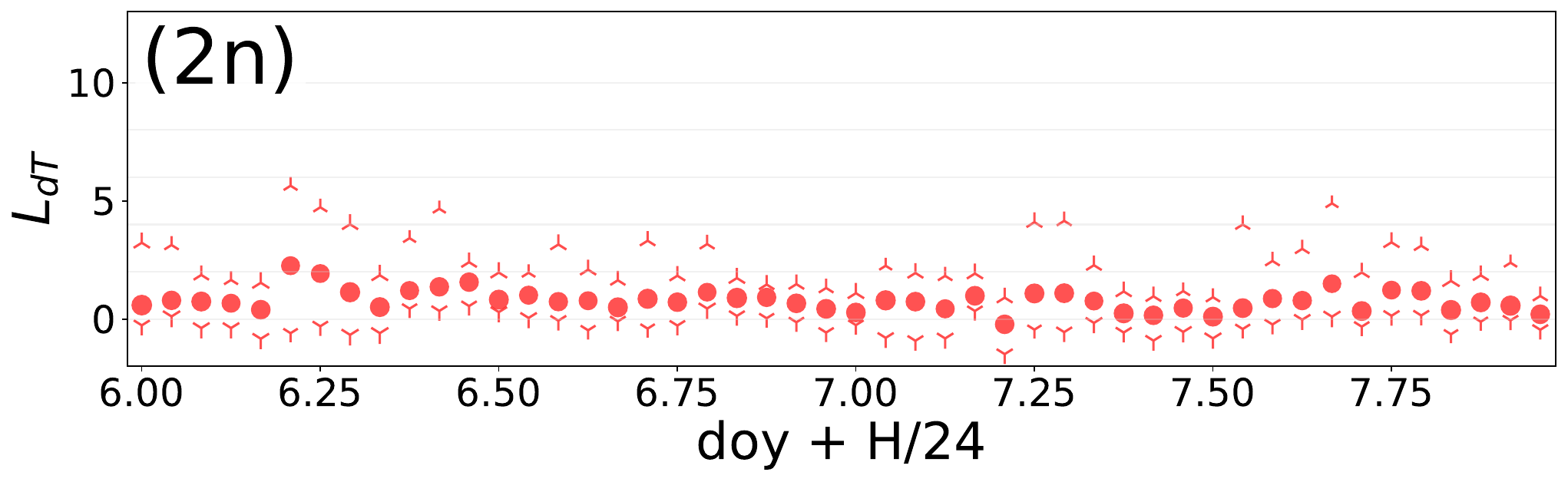}
\includegraphics[width=0.99\columnwidth]{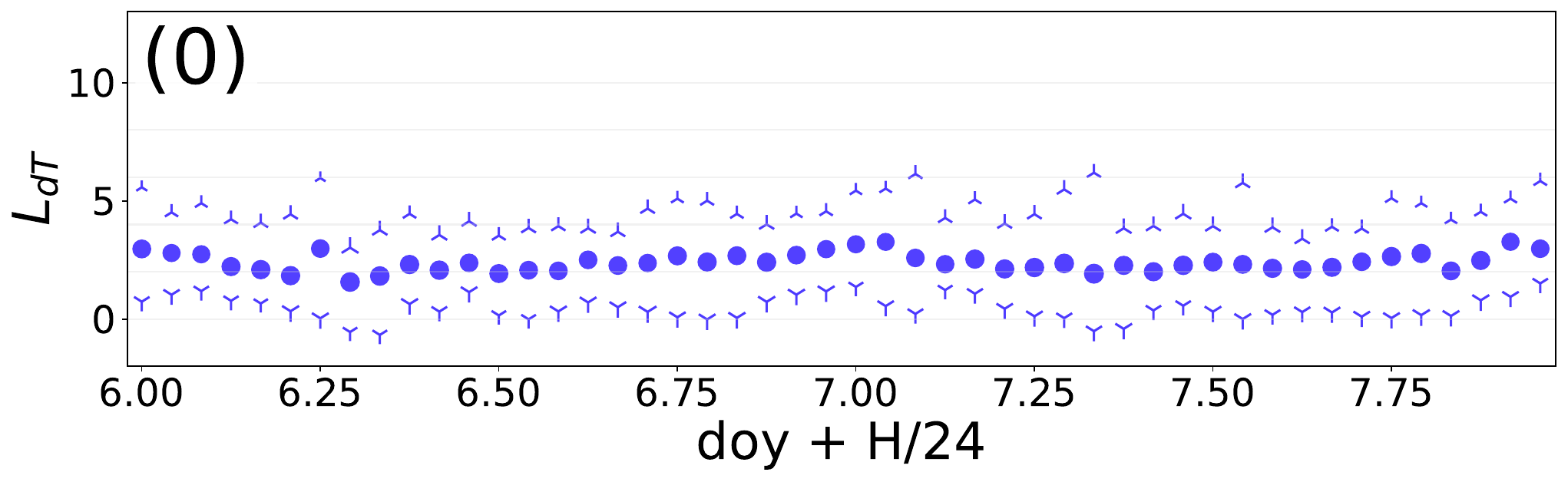}
\includegraphics[width=0.99\columnwidth]{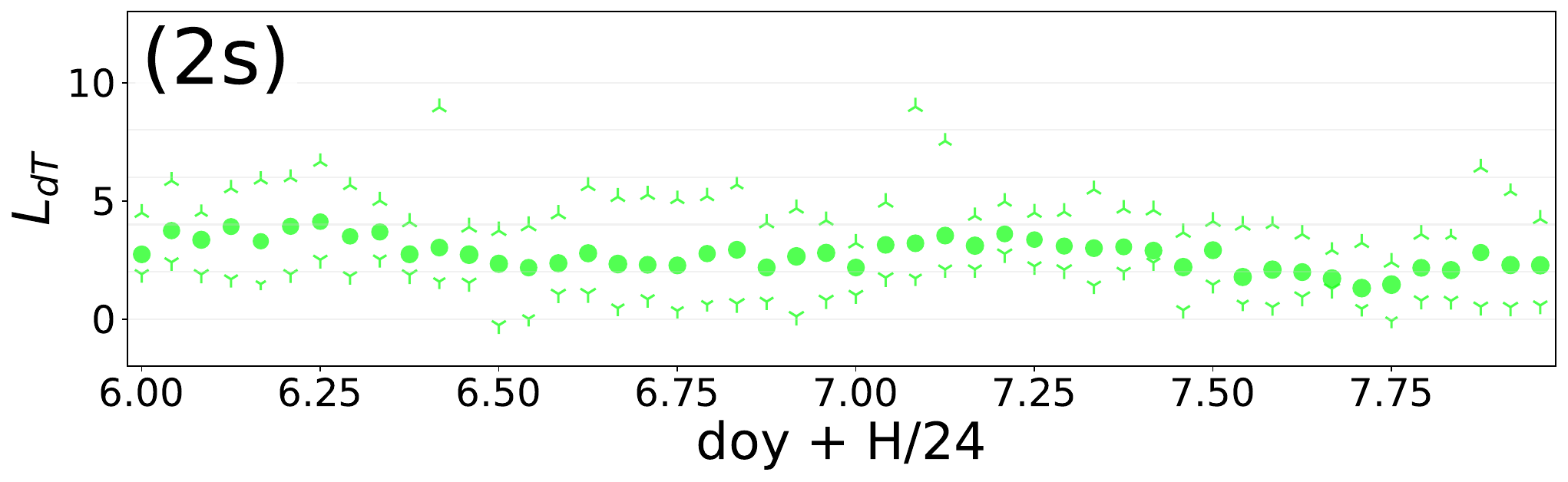}
\caption{Log scale index $\LdT$
 for our chosen quiet days in 2022 006/007;
 showing
 variation by time (UT) and longitude dodecant for 
 high northern magnetic latitudes (2n), 
 low latitudes (0), 
 and high southern magnetic latitudes (2s).
Here we indicate the range in $\LdT$ activity by
 dividing the dodecant $\LdT$ widths up into three, 
 the two maximum  (averaged to give points $\curlywedge$), 
 the two minimum  (averaged to give points $\curlyvee$), 
 and the remaining eight widths (averaged to give points \textbullet).
}
\label{fig-LdT-2022-00x}
\end{figure}

\begin{figure}[ht]
\includegraphics[width=0.99\columnwidth]{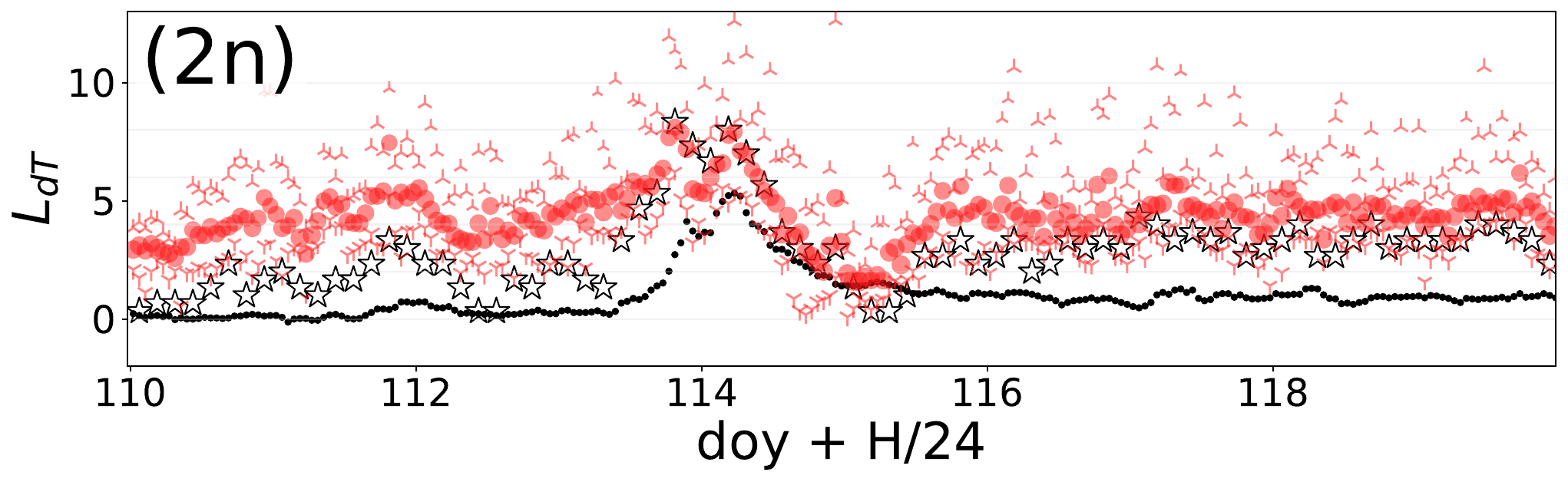}
\includegraphics[width=0.99\columnwidth]{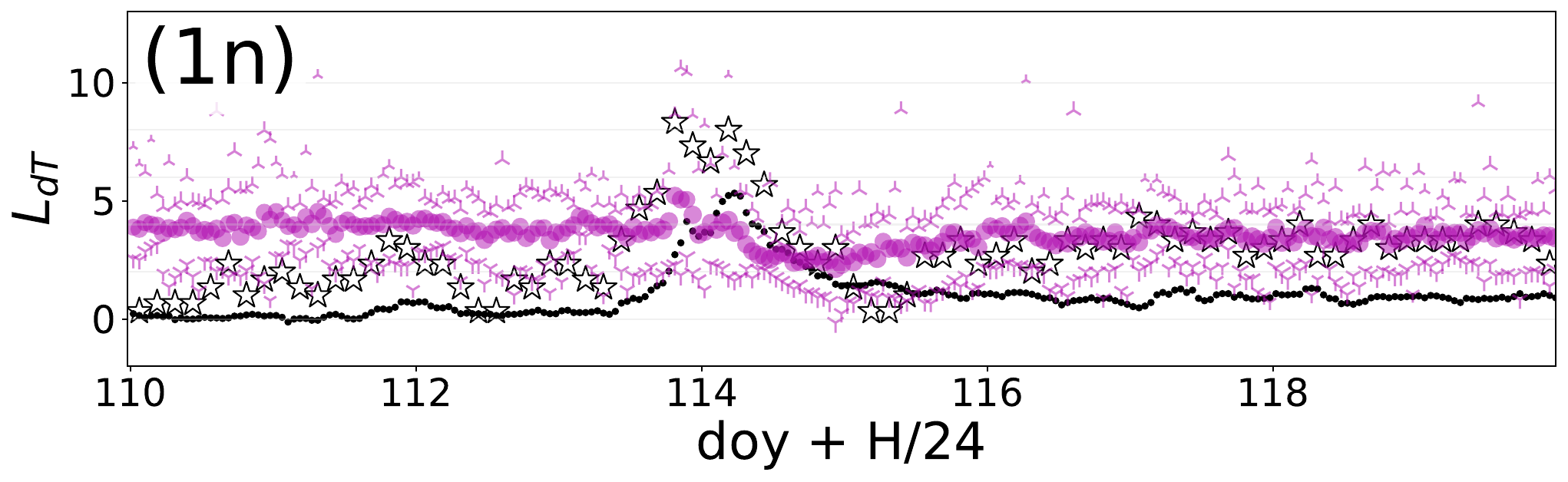}
\includegraphics[width=0.99\columnwidth]{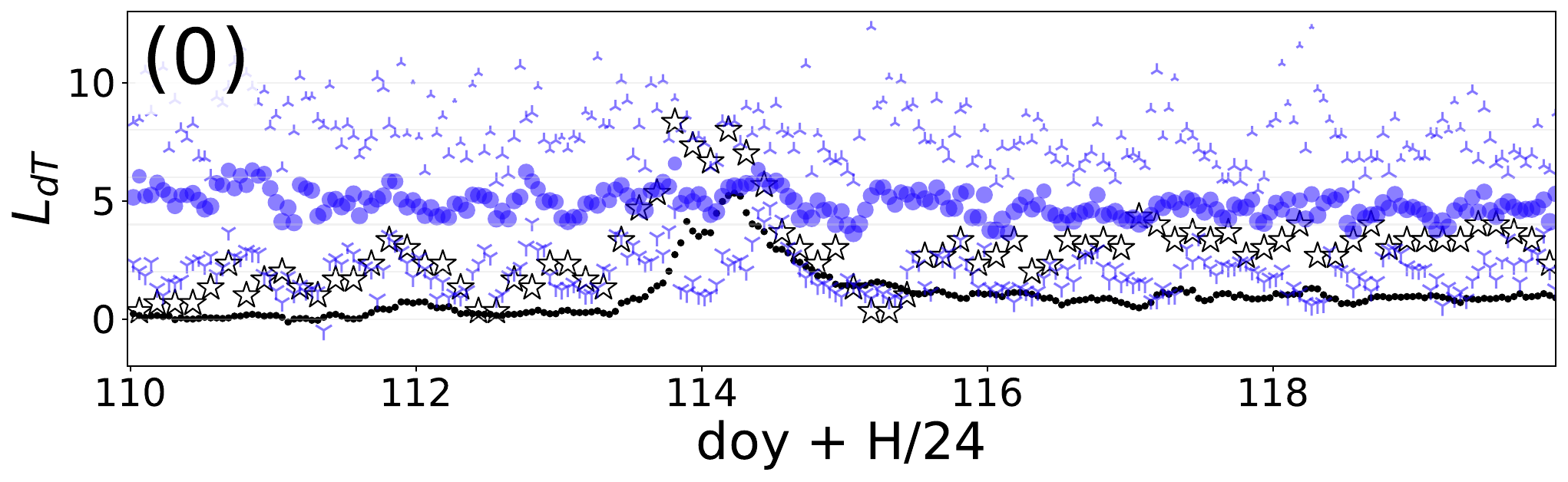}
\includegraphics[width=0.99\columnwidth]{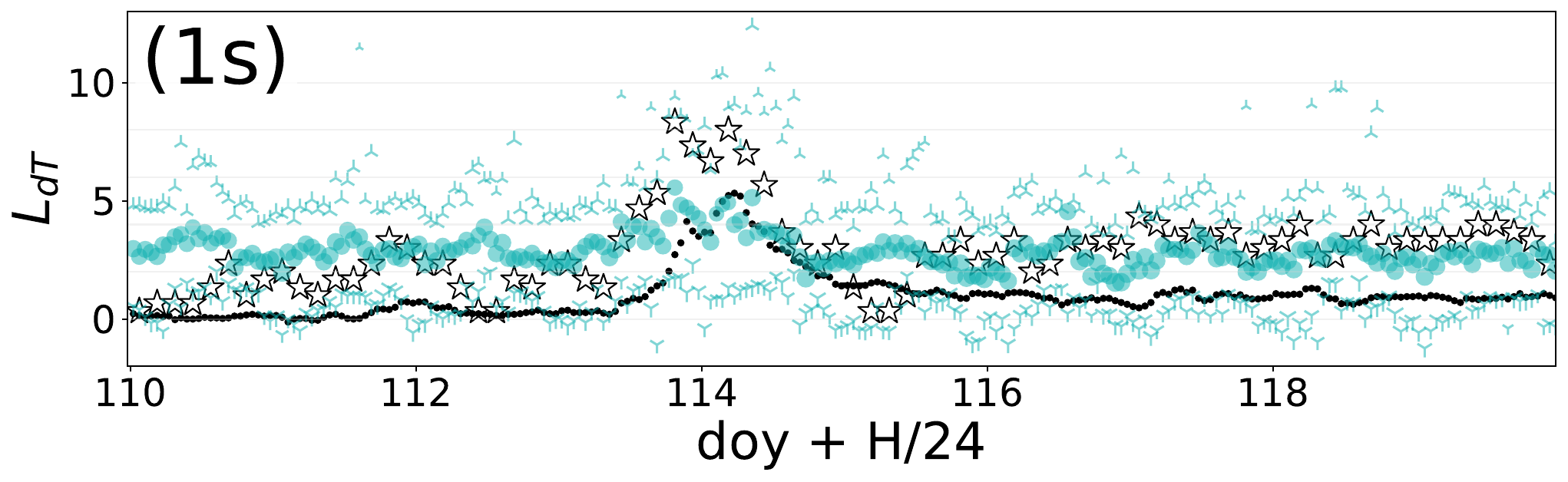}
\includegraphics[width=0.99\columnwidth]{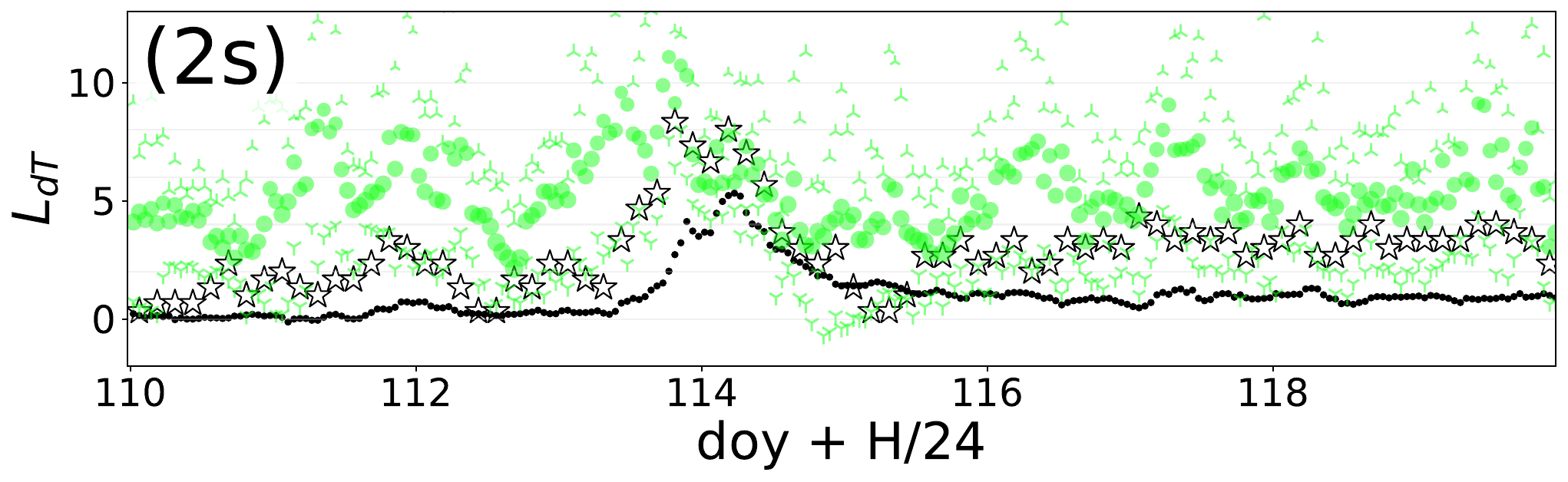}
\caption{Log scale index $\LdT$
 for our chosen active days in 2023, 110 to 119;
 showing -- as done in Fig. \ref{fig-LdT-2022-00x} --
 variation by longitude dodecant for 
 high and mid northern magnetic latitudes (2n, 1n), 
 low latitudes (0), 
 and mid and high southern magnetic latitudes (1s, 2s).
Here we indicate the range in $\LdT$ activity by
 dividing the dodecant $\LdT$ widths up into three, 
 the two maximum  (averaged to give points $\curlywedge$), 
 the two minimum  (averaged to give points $\curlyvee$), 
 and the remaining eight widths  (averaged to give coloured filled circles).
The variation in the three-hourly $\KPI$ index
 over its standard range from 0--9
 is also indicated on each as the open star {\FiveStarOpen}.
The contemporaneous hourly variation in $\DST$ index
 is indicated by black dots, 
 where the value plotted is $-\DST/40$, 
 so that it also fits within the 
 vertical axis range of 0--10.
Note that although the $\KPI$ and $\DST$ behaviour 
 might be said to match reasonably well with our $\LdT$
 index for high northern latitudes, 
 it does not do so elsewhere.
}
\label{fig-LdT-2023-11x}
\end{figure}

On Figs. \ref{fig-LdT-2022-00x} and \ref{fig-LdT-2023-11x}
 we can see how this $\LdT$ index varies over time hour-by-hour,
 and how it varies between longitude dodecants for high and low magnetic latitudes.
Note that although plotted using this logarithmic index, 
 these figures would, 
 at a glance,
 look rather similar if the
 $\WdT$ width measure had been plotted instead.

\newcommand{\hollowstar}{\ding{80}}

\begin{figure}[ht]
\includegraphics[width=0.99\columnwidth]{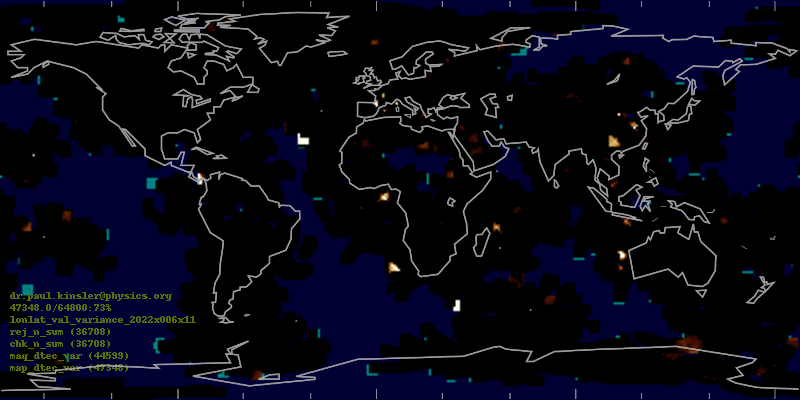}
\caption{``Quiet day'' activity map: this depiction
 is obtained by sorting the moderated dTEC values into overlapping geographic 5x5 degree pixels and computing 
 their variance. This map shows a not-untypical hour (11:00UTC) from the quiet day
2022/006; because it is quiet the variances are typically small
 so the map is almost uniformly dark --
 contrast this with Figure \ref{fig-LdT-2023-11x-map}.
Here, dark blue
 areas indicate no data, 
 otherwise the color scale ranges from black (zero variance) 
 through red, then to white for variances $>=0.56$.   
Fitting distributions in these pixels (as per Section \ref{S-fitting}) is not done because
 in general there are insufficient dTEC values per pixel.
The isolated bright white squares that sometimes
appear on these automatically generated maps are typically due to
difficult to filter data anomalies or processing artifacts.
}
\label{fig-LdT-2022-00x-map}
\end{figure}


\begin{figure}[ht]
\includegraphics[width=0.9799\columnwidth]{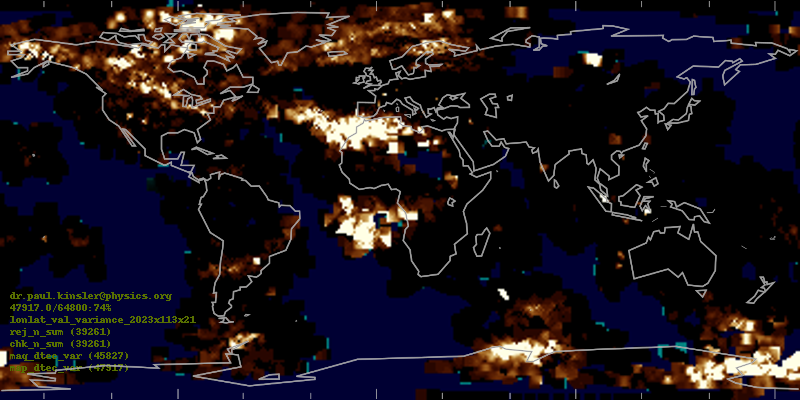}
\includegraphics[width=0.9799\columnwidth]{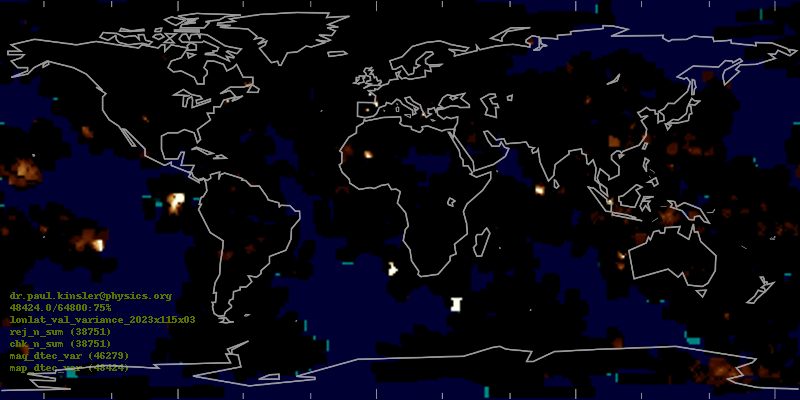}
\caption{``Active day'' maps: these depictions
 show moderated dTEC variances
 accumulated for 
 \cRed{a very active hour during the April 2023 storm,}
 i.e. 21:00UTC on 2023/113 (top);
 and 
 the subsequent least active minimum at 03:00UTC on 2023/115 (bottom). 
These are plotted in the same manner, 
 and with the same plotting range,
 as in Fig \ref{fig-LdT-2022-00x-map}.
}
\label{fig-LdT-2023-11x-map}
\end{figure}


\begin{figure}[ht]
\includegraphics[width=0.99\columnwidth]{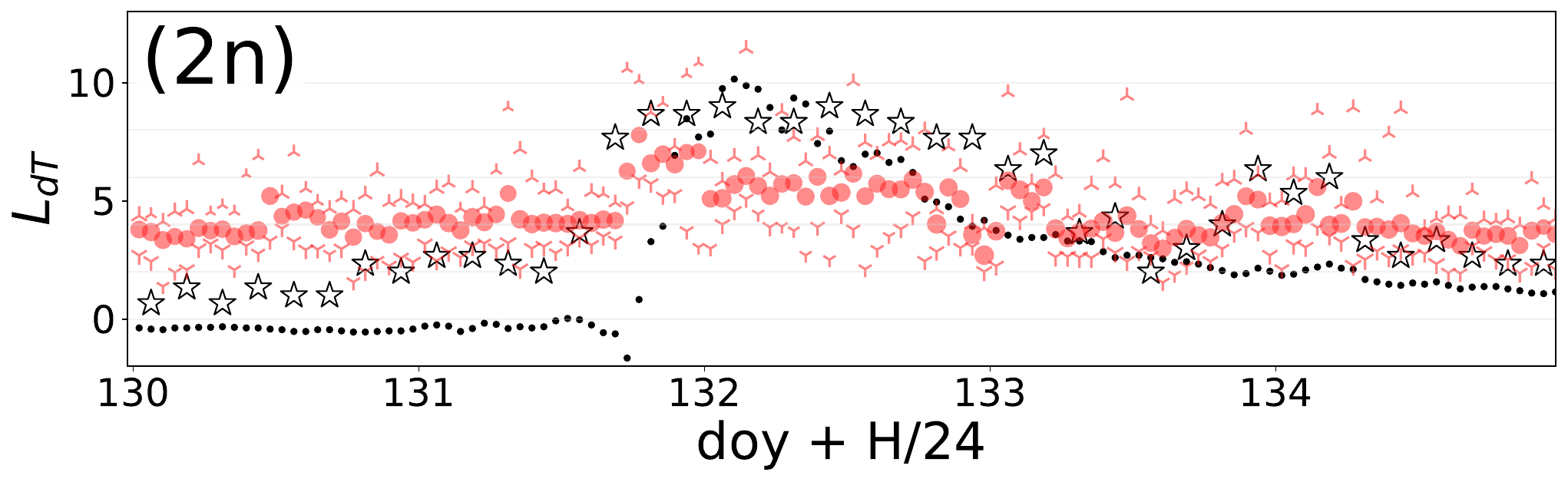}
\includegraphics[width=0.99\columnwidth]{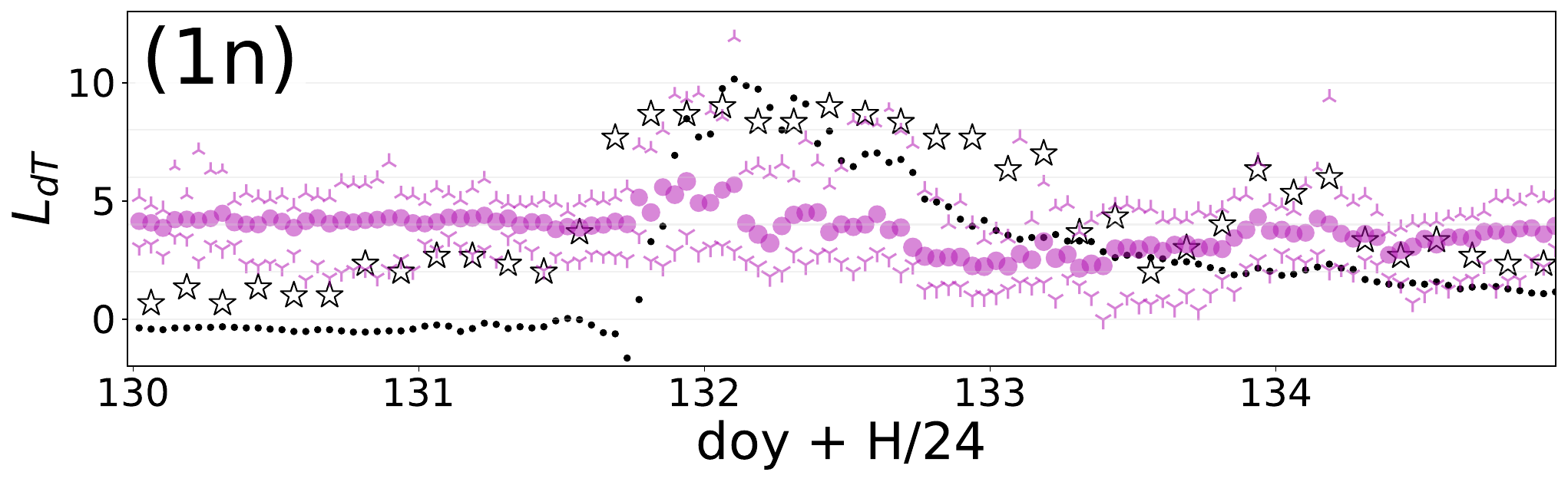}
\includegraphics[width=0.99\columnwidth]{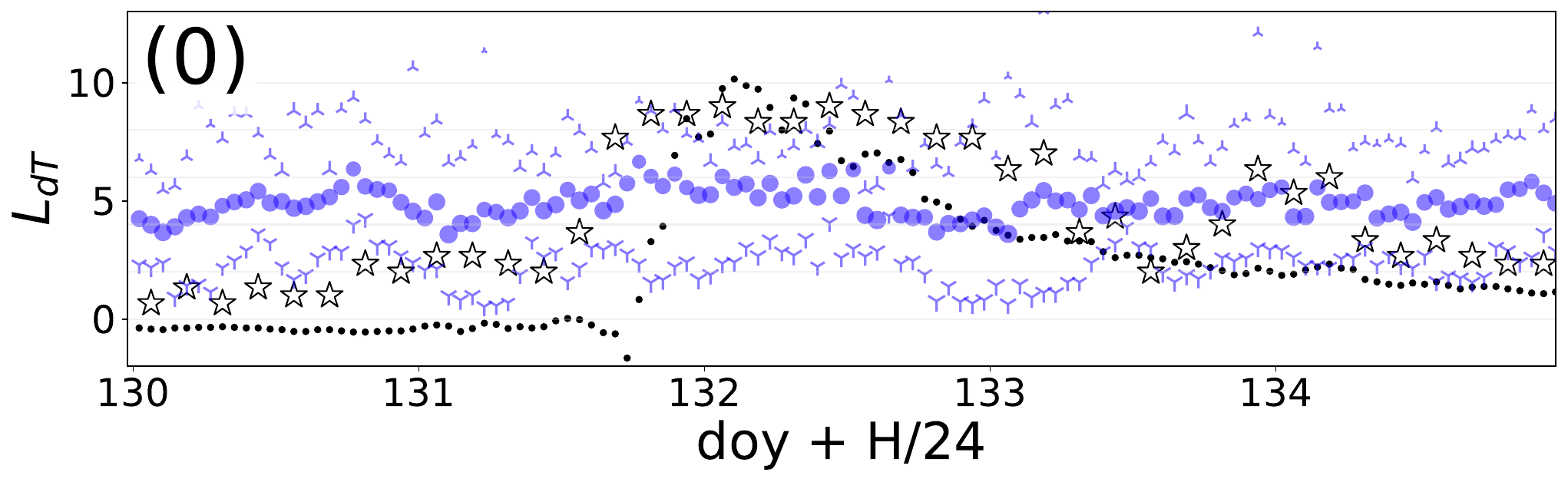}
\caption{Log scale index $\LdT$ for high (2n) and mid (1n) northern latitudes, 
 and equatorial (0) latitudes,
 for 2024's May storm event 2024 over doys 130 to 134; 
 plotted as per Fig. \ref{fig-LdT-2022-00x}.
The variation in the $\KPI$ index
 is also indicated on each as the open star {\FiveStarOpen};
 and that in $\DST$ index
 is indicated by black dots plotted at values $-\DST/40$.
Note that unlike the 2023 April storm days 
 there is significantly poorer agreement between $\KPI$, $\DST$, and $\LdT$.
}
\label{fig-LdT-2023-13x}
\end{figure}

For the quiet day result shown on Fig. \ref{fig-LdT-2022-00x},
 notice that the variation in all ``remainder'' averaged $\LdT$'s
 is only a factor of about two,
 and have a maximum $\LdT$ of about four;  
 and that the spread between maximum/minimum markers  
 is usually relatively contained.
Notwithstanding this general quiet behaviour, 
 we can still see some outliers where $\LdT$ was high, 
 notably at southern high latitudes (2s).
The low activity levels seen on the map 
 Fig. \ref{fig-LdT-2022-00x-map}
 generated for 11:00UTC on doy 006 are in agreement with the 
 $\LdT$ measures on Fig. \ref{fig-LdT-2022-00x}.
Of course, 
 more detail could be brought out by altering the colour scale, 
 but we retain our default so that this map can be compared 
 directly to those with strong activity, 
 as are shown later in Fig. \ref{fig-LdT-2023-11x-map}.

In stark contrast to those quiet day results, 
 for the April 2023 active days shown on Fig. \ref{fig-LdT-2023-11x}, 
 notice that on day 114, 
 even the average (``remainder'') $\LdT$ for high northern latitudes (2n) 
 decreases from about 8 down to 2, 
 at high southern latitudes (2s) the trend is noisier
 but nevertheless varies from 10 down to 4 -- 
 i.e. both by a factor of about eight ($2^3$); 
 whereas the change at equatorial latitudes is much smaller
 (about 2).
Further, 
 the maximum/minimum markers indicate considerable variation 
 between different longitude dodecants.
In Fig. \ref{fig-LdT-2022-00x-map}
 we show geographical activity maps
 timed at the apparent northern polar peak (21:00UTC doy 113)
 and then in its 
 trailing dip at 03:00 on doy 115; 
 these both show the widespread extent of GNSS-effective activity 
 during those storms.

However, 
 although on Fig. \ref{fig-LdT-2023-11x}
 there looks to be a remarkable match between variation of $\KPI$
 and the $\LdT$ for high northern latitudes (2n), 
 note that here $\KPI$ does not provide a good indicator
 in other latitude bands --
 not even for high southern latitudes.
Further,  
 on Fig. \ref{fig-LdT-2023-13x} we show a similar comparison 
 using recent data for the May 2024 storm, 
 and in this case -- as shown at at high northern latitudes --
 we can see that $\KPI$ seems less convincing as a proxy 
 for the GNSS-effective activity indicated by $\LdT$.
In addition, 
 and arguably more clearly than in the April 2023 data, 
 we can see here in May 2024 how ionospheric disturbances 
 expanded equatorwards into middle latitudes
 at North American longitudes.
This comparison supports the need for a scale
 that is capable of describing the state of the ionosphere
 in response to specific space weather conditions,
 and that can be linked to an impact.
The $\LdT$ index describes the state of the ionosphere
 by utilising information from dTEC
 (hence, disturbances in radio propagation) 
 as observed through GNSS links.

As a final note,
 the $\LdT$ data presented in this section 
 is based on the dTEC distribution widths $\WdT$, 
 and so does not incorporate the power law tail components
 added to $\WdTs$.
As already discussed, 
 the relative importance of these tails is uncertain, 
 so we chose to omit their effects here,
 and leave any detailed analysis to later work.

%
\section{Conclusion}\label{S-conclusion}

An important aspect of the ionospheric plasma behaviour
 is the occurrence of irregularities in the electron density distribution
 that form in conjunction of plasma instability mechanisms
 initiated under specific space weather conditions. 
Propagation disturbances associated with ionospheric irregularities
 have an immediate consequence for application and services
 reliant upon satellite radio signals that propagate through the ionosphere
 (e.g. GNSS).
It is the occurrence of propagation disturbances
 (e.g. phase fluctuations and scintillation)
 that can adversely impact applications reliant upon
 satellite positioning,
 navigation,
 and timing,
 as a result of the degradation in positioning quality
 \cite{Fabbro-JLR-2021jswsc,John-FAAAVHS-2021rsc,Forte-AAAVMSKJ-2024asr,
Nykiel-CHJ-2024gpss,Roberts-FJ-2019gpss}.

Here we introduced a method to describe the spatial variation
 and temporal evolution of ionospheric irregularities on a global scale,
 based on the temporal rate of change of TEC (dTEC).
Notably,  
 we based our analysis on the statistical distribution of dTEC values, 
 a step which enables us to cope easily with their power-law tails, 
 an improvement over (e.g.) reducing the data to a simple variance.

The first part of the method starts with uncalibrated slant TEC
 and its rate of change by using geometry-free combinations
 of the carrier phases
 from all the dual-frequency combinations available at each ground station. 
To extract key characteristics from a available data slices, 
 we binned and fitted frequency histograms
 summarizing computed temporal changes in dTEC values; 
 where the fit process enable us to not only avoid problems with potentially unbounded moments, 
 but also characterising the power-law tails
 that make moments problematic.
Our method and metric differs from existing proposals such as ROTI maps,
 as it calculates the variability of dTEC
 from all available links intersecting a given ionospheric pixel over any given hour,
 instead of utilising the average of link-specific ROTI values 
 \cite{Jakowski-SSK-2006asr,Jakowski-BW-2012rsc,
Cherniak-KZ-2014rsc,Cherniak-Z-2016eps,Cherniak-KZ-2018eoi,Wilken-KJB-2018jswsc,
Jakowski-H-2019sw,Denardini-PBNCRMCRSB-2020sw,
Kotulak-ZKCWF-2020rse,John-FAAAVHS-2021rsc,
Fabbro-JLR-2021jswsc,Kotulak-KFFWLB-2021sen,Nykiel-CHJ-2024gpss}.
Our approach 
 is sensitive to the presence of large-scale ionospheric irregularities, 
 and it informs on 
 the spatial and temporal changes in ionospheric properties.
Some of these changes are part of a relatively regular or seasonal behaviour -- 
 such as the higher occurrence of irregularities and scintillation
 in the equinoctial post-sunset equatorial ionosphere,
 but some are due to external solar-origin space weather forcing.
Both types are captured equally well by our analysis.

The second part of the method estimates an ionospheric scale $\LdT$
 that summarises and encapsulates the state of the ionosphere
 under varying solar and geomagnetic conditions. 
In contrast to indices such as $\KPI$ and $\DST$,
 our ionospheric scale can be --
 and is -- 
 easily estimated for different regions,
 so that it can retain sensitivity to spatial and temporal dependencies
 such as in the case of geoeffective active conditions. 
Further, 
 since our index is defined in terms of a process
 and is not tied to any specific dataset, 
 there is no impediment to defining new regions or sub-regions
 of particular interest, 
 or using either UTC ordering or local time ordering, 
 and computing an $\LdT$ index as needed
 from whatever set of ground station data that is most applicable.

Our method was tested on quiet and active case studies,
 and it demonstrates the ability to distinguish between day-to-day variability,
 more persistent patterns, and the response to storm conditions. 
The proposed ionospheric scale could be utilised by service providers
 aimed at conveying information in an intuitive fashion, 
 informing about the state of the ionosphere, 
 for example, in relation to possible impacts on applications.

%
\acknowledgments

The work at the University of Bath was supported by
 the UK Natural Environment Research Council
 (Grant number NE/V002597/1 and Grant number NE/X019004/1).
The $\KPI$ index data were obtained from the GFZ International $\KPI$ index Service
 (ftp://ftp.gfz-potsdam.de/pub/home/obs/); 
$\DST$ data were obtained from the WDC for Geomagnetism service 
 (https://wdc.kugi.kyoto-u.ac.jp);
 and 
 RINEX data were accessed through the
 International GNSS Service (IGS)
 (https://cddis.nasa.gov/archive/gnss/products/)
 \cite{CDDIS-NASA,Johnston-RH-2017shignss}.

%
%

%
\clearpage

\appendix*

\setcounter{page}{1}
\pagenumbering{roman}

\section{Supplementary Material: \\
$\LdT$: An indicator of ionospheric activity
 based on
 statistical distributions
 in GNSS-derived TEC rates of change}

\begin{centering}
  {\emph{Paul Kinsler, Biagio Forte}\\
~

{Department of Electronic and Electrical Engineering
  University of Bath, Bath BA2 7AY,
  United Kingdom}
}

\end{centering}

~\\
~\\

\section{A. Code and data processing}\label{S-ionwork}

\def\IONW{IONwork}

The code is a combination of (bash) shell scripts and python3 code 
 with minimal dependencies.
The bash scripts are mainly used for workflow management -
 i.e. supervising the running of the python code with suitable arguments
 and in the right order -
 but also is used with (e.g.) unix stream editing tools to extract basics
 like satellite and ground station names,
 and data timestamps from data and configuration files. 
The python codes generally handle the more sophisticated operations,
 making heavy use of its list comprehension abilities
 to filter and sort large lists to find and step-through appropriate subsets. 
Output data is saved in a custom (but fairly generic) space-separated text format,
 so that conversion of output into other formats 
 is simple.

The main processing steps proceed as follows:

%
\subsection{Downloads}

RINEX files containing 30 second interval data 
 are downloaded as hourly or daily sets
 from CDDIS \cite{CDDIS-NASA}.
This typically provides data from over 300 ground stations
 distributed around the world.
However, 
 RINEX data from other sources
 can also be processed.

%
\subsection{File conversion}

To simplify processing we convert the downloaded RINEX files
 into simpler (but less compact) lists, 
 and extract some summary information 
 such as lists of reporting ground station, 
 their locations, 
 the satellites that were seen, 
 and so on.
The code can automatically extract all observation (receive) times
 from the data, 
 but usually we use the 30 second observation interval
 to directly construct a list of all possible observation times.
A summary of GLONASS sideband usage is also compiled, 
 as precise frequencies are needed later for cycle-slip detection.

Some of these conversions fail due to corrupted RINEX data, 
 so the code is written to be as tolerant as possible 
 of such complications.

%
\subsection{Satellite trajectory prediction}

Given the list of observation times, 
 we calculate each seen satellite's position at that time. 
Since this is a complicated processs, 
 we have checked the positions predicted by our codes
 against those computed by gLAB \cite{gLAB-2018} from the same navigation data.
For GPS, 
 GALILEO, 
 Beidou, 
 and GLONASS satellites, 
 we see negligible differences when 
 predictions are compared.

%
\subsection{Ionosphere intersections}

Here we assume a spherical Earth, 
 and a thin shell-like ionosphere at 350km.
Since the spatial and temporal resolution eventually used
 is rather coarse, 
 any discrepancies due to these simplifying assumptions are negligible.

With these assumptions, 
 the computation of the line of sight (LOS) intersections between each possible
 satellite and ground station pair is a simple exercise in geometry.
To ensure we can later correct for the LOS angle as it intersectes the ionosphere, 
 as well as keep track of any horizon-skimming recieve events, 
 we record its cosine in the resulting data files. 

%
\subsection{Observations}

Matching up the LOS intersections with the received data
 is simply a process of correlating the LOS markers of time, satellite, and station 
 with those in the converted RINEX observation files.
This step is, 
 however, 
 one of the more computationally demanding ones.

%
\subsection{Checking}

Since we intend to compute and use TEC values from these observations, 
 we also filter the data to remove cycle slips
 with an implementation of the Melbourne-Wubbena algorithm.
In addition, 
 we also remove short data lines (since we need at least two matched code-signals
 in different frequency bands),
 and isolated detections (since we are only interested in differences).

The rejected observation lines are stored separately from the acceptable ones, 
 and can be use to e.g. plot or tabulate the number of cycle slip events.

%
\subsection{TEC computations}

Here we compute the slant TEC from every possible pair of code-matched 
 signals in different frequency bands, 
 i.e. not just L1C and L2C but all of the thirty or so possibilities, 
 which are listed on Table \ref{table-lpnorms-full}.
This simply requires measured phase data by signal band and coding,
 signal frequencies,
 and the the standard formula
 from \eqref{eqn-TECestimate}.

%
\subsection{dTEC values (``events'')}

Having computed the slant TEC, 
 we then follow each station-satellite pair and 
 use the TEC's change in time over the 30s observation interval
 to compute its time derivative (dTEC) in TEC units-per-second.
Note that 
 since satellites are moving above the Earth's surface,
 no two TEC estimates ever correspond to exactly the same intersection point
 with the ionosphere, 
 but (as above) since our analysis is coarse-grained in space --
 not going beneath 1-by-1 degree pixels -- 
 this motion does not greatly affect our analysis.
However,  
 the resulting motion does
 impinge on the size of ionosperic features we might resolve, 
 and on how we might detect their temporal behaviour or motion.
Notably, 
 since the TEC estimates are not strictly comparable because
 they result from different signal paths,
 by comparing them we are
 neither probing solely temporal properties
 nor solely spatial properties of any ionospheric disturbances, 
 but a combination of both.

%
\subsection{dTEC digests}\label{S-ionwork-digests}

Before creating our digests
 (histogram summaries)
 we filter and rescale the raw dTEC values
 in a variety of ways.

First, 
 we usually filter the dTEC events 
 so as to avoid horizon and/or multipath complications
 due to signal receives made at low angles.
However, 
 rather than use an angle-based ad hoc rule such as
``ignore all receives less than 20 degrees above the horizontal''
 we looked at the dTEC values as a function of angle and
 made a cut at the point where error rates
 (cycle slips, drop outs, etc)
 rose above about third; but 
 whose exact value was chosen so the the cut-off
 conveniently lies at an intersection cosine with the ionosphere of exactly 0.4. 
As it happens, 
 our chosen cutoff is similar to the 20 degree one, 
 although we had not set that as a criteria.

Second, 
 as discussed in Sec. \ref{S-norm-pathangle},
 we corrected the 
 dTEC values for the geometry of their LOS
 path through the ionosphere
 according to our data-driven proceedure.

Third,
 and as discussed in Sec. \ref{S-norm-lpair},
 we normalised dTEC values based on band pair, 
 according to the different properties (widths) 
 seen for their respective dTEC distributions
 for our selected quiet days (2022/006 and 007), 
 \cRed{as well as 215/328 and 329, 
 and 2020/314 and 315.
Each pairing was normalised separately, 
 then the $B$ for each band pair
 was picked according to the best-sampled case.}
A full table of scalings is given in Table \ref{table-lpnorms-full}.

\begin{table}
\begin{center}
\begin{tabular}{ || l | c || }
\hline
\hline
Band Pair  & $B(1.60)$  \\
\hline
\hline
L1CL2C & 1.0000 \\
L1CL6C & 1.1482 \\
L1DL5D & 1.4421 \\
L1DL7D & 1.4452 \\
L1LL2L & 1.0267 \\
L1PL2P & 1.0174 \\
L1PL5P & 1.2627 \\
L1WL2W & 1.2596 \\
L1XL2X & 0.9592 \\
L1XL5X & 1.1766 \\
L1XL6X & 1.1320 \\
L1XL7X & 1.2355 \\
L1XL8X & 1.2732 \\
L2IL6I & 1.1848 \\
L2IL7I & 1.4405 \\
L2XL5X & 0.4862 \\
L2XL6X & 1.3340 \\
L2XL7X & 1.4314 \\
L5DL7D & 0.4798 \\
L5QL7Q & 0.4479 \\
L5QL8Q & 0.3839 \\
L5XL6X & 0.7855 \\
L5XL7X & 0.3607 \\
L5XL8X & 0.2823 \\
L6IL7I & 0.5645 \\
L6XL7X & 0.6460 \\
L6XL8X & 0.7566 \\
L7QL8Q & 0.4193 \\
L7XL8X & 0.2844 \\
\hline
L1CL2W & 1.1770 \\
L1CL2X & 0.9715 \\
\hline
L1cL2c & 1.1055 \\
L1cL5c & 1.0519 \\
L1cL6c & 0.4334 \\
L1cL7c & 1.2735 \\
L1cL8c & 1.2438 \\
L2cL5c & 0.6355 \\
L5cL6c & 1.0992 \\
L5cL7c & 0.6302 \\
L5cL8c & 0.4580 \\
L6cL7c & 0.9175 \\
L6cL8c & 1.0644 \\
L7cL8c & 0.4766 \\
\hline
\hline
\end{tabular}
\end{center}
\caption{Computed dTEC normalisation scalings for the 
 30 band-pairs used to calculate TEC and dTEC values, 
 as based on aggregated data from the ``quiet'' days 006 and 007 of 2022,
 \cRed{as well as 215/328 and 329, 
 and 2020/314 and 315.}
The dTEC values for each individual band-pair
 were scaled using the path-angle correction $C^\eta$,
 then combined into a histogram/distribution
 the G2E fitting applied, 
 and a G2E weighted width calculated.
This was then taken as a ratio against the width 
 \cRed{for the L1CL2C bandpair.}}
\label{table-lpnorms-full}
\end{table}

\section{B: Distributions and fits}\label{S-distribufit}

\begin{figure}
\includegraphics[width=0.90\columnwidth]{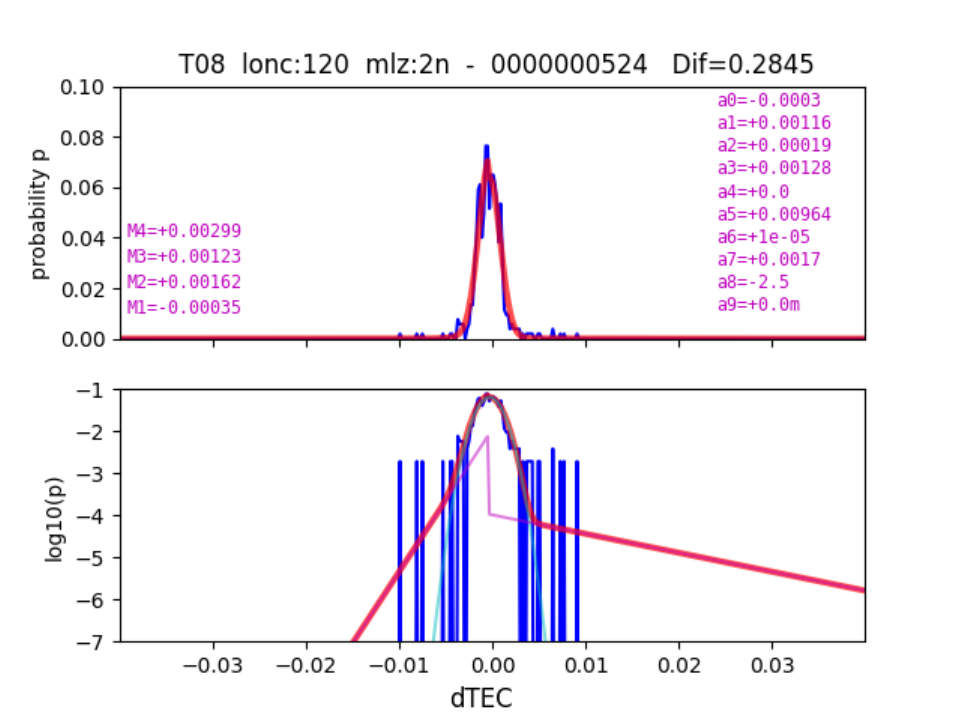}
\includegraphics[width=0.90\columnwidth]{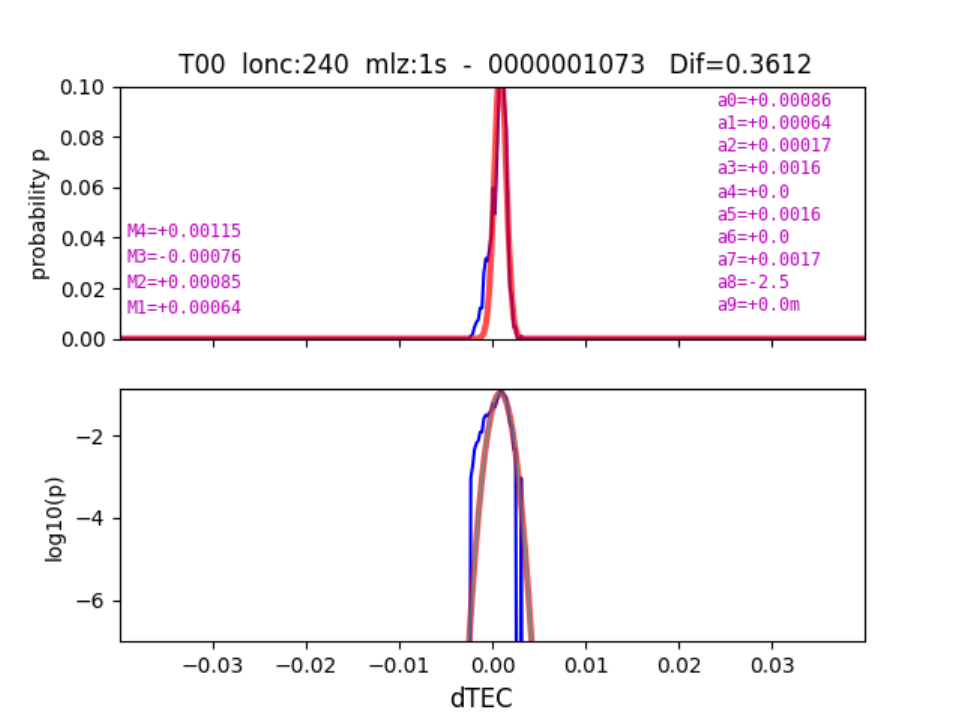}
\caption{Some low event count (less than 1k) 
 event histograms and fits from the quiet day 007 in 2022; 
 where the panel titles indicate the specific data slice addressed, 
 the ``M'' parameters on the left are moments, 
 and the ``a'' on the right are the fitted $\alpha_i$.
These distribution shapes are not untypical,
 but these two by no means indicate the true variation present.
In these figures the blue lines represent
 the data-derived pDF for dTEC values, 
 whereas the red line is the fit. 
In most log plots you can also see a cyan line, 
 which indicates the gaussian contribution to the overall fit; 
 likewise the magenta line indicates the exponential contribution.
\cRed{In the top ``T08'' panel
 (at lonc:120 and mlz:2n; and based on 284 dTEC values)
 we see both gaussian and exponential contrbutions, 
 and a significant asymmetry; 
 whereas 
 in the bottom ``T00'' panel (lonc:240, mlz:1s, 1073 values)
 the fit is mainly gaussian.
However, 
 in these two low-sampling cases,
 out ``Dif'' fitting errors are relatively large, 
 at 0.2845 and 0.3612 respectively.}
}
\label{fig-histo1E2}
\end{figure}

\begin{figure}
\includegraphics[width=0.90\columnwidth]{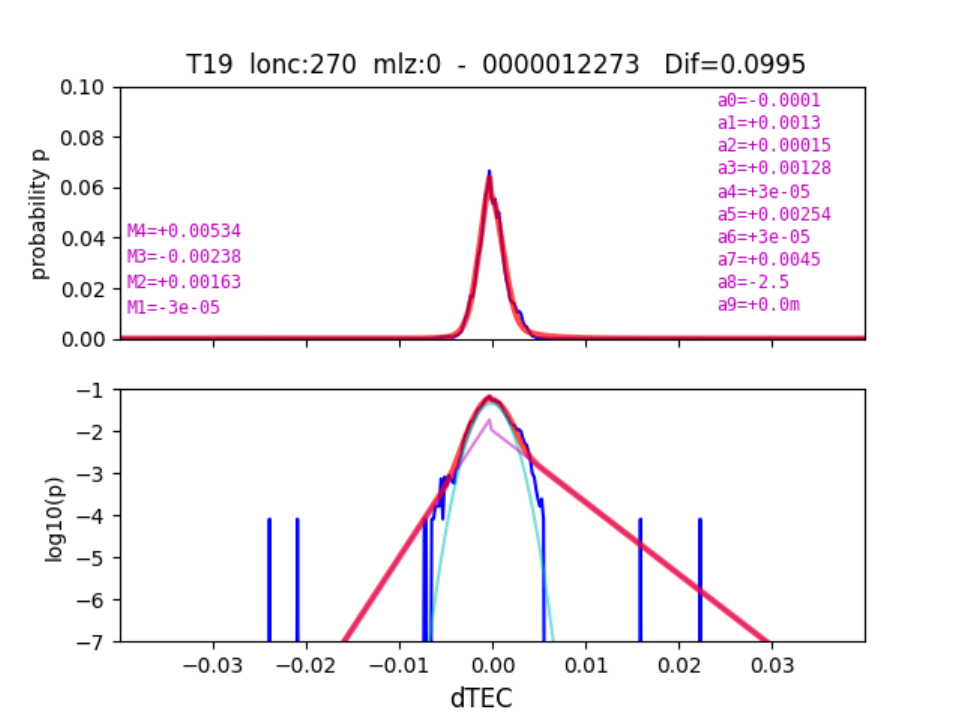}
\includegraphics[width=0.9800\columnwidth]{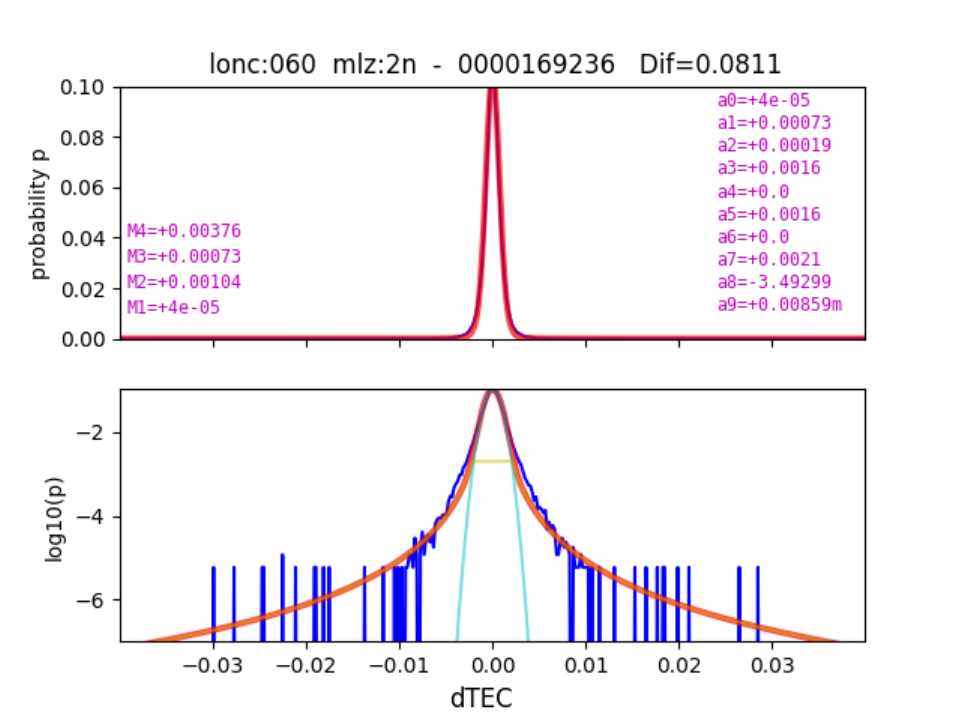}
\includegraphics[width=0.9800\columnwidth]{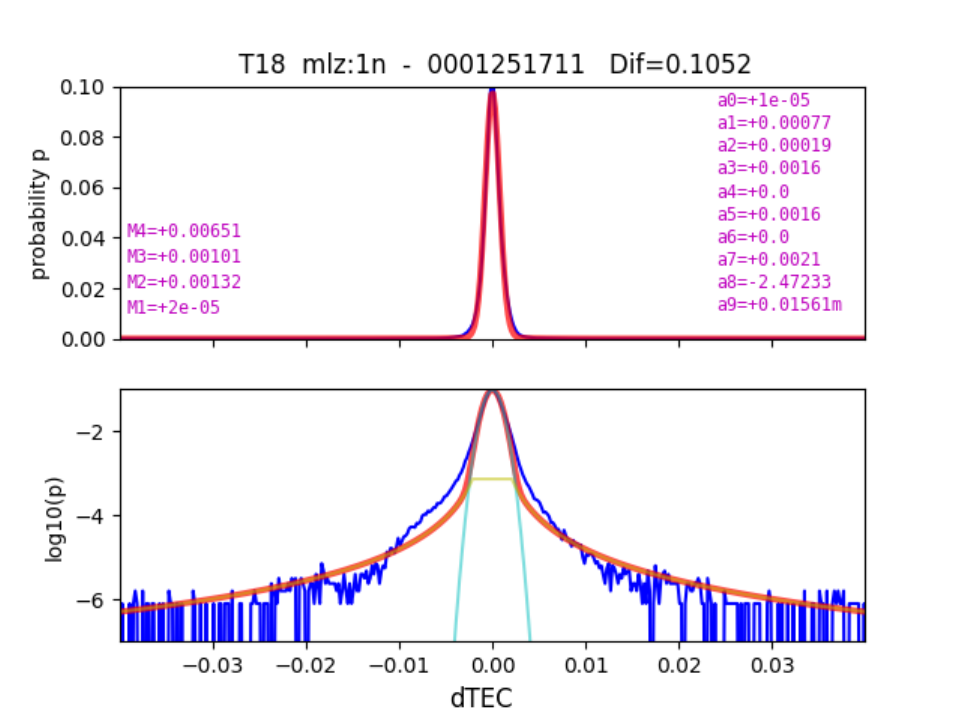}
\caption{\cRed{Increasingly aggregated distributions, 
 for more populated data slices from 2022/007, 
 containing nearly 12k, 169k, and 1.2 million dTEC values respectively.
Again,
 these distribution shapes are not untypical, 
 but neither do they represent the full variation.
In the upper ``T19'' panel based on 12273 dTEC values, 
 the exponential contribution helps
 match the central core to inner-wings better, 
 but not the wider wings; 
 and in the middle ``lonc:060'' panel (from 169236 values) we can 
 see that no exponential behaviour was detected, 
 but that there is a power law tail in evidence.
The lower ``T18'' panel (1251711 values) again 
 has a gaussian core with no exponential part, 
 and a much better sampled power law tail appears.}
Line colours \& etc are the same as for Fig. \ref{fig-histo1E2}.
}
\label{fig-histo1E4}
\end{figure}

\begin{figure}
\includegraphics[width=0.9800\columnwidth]{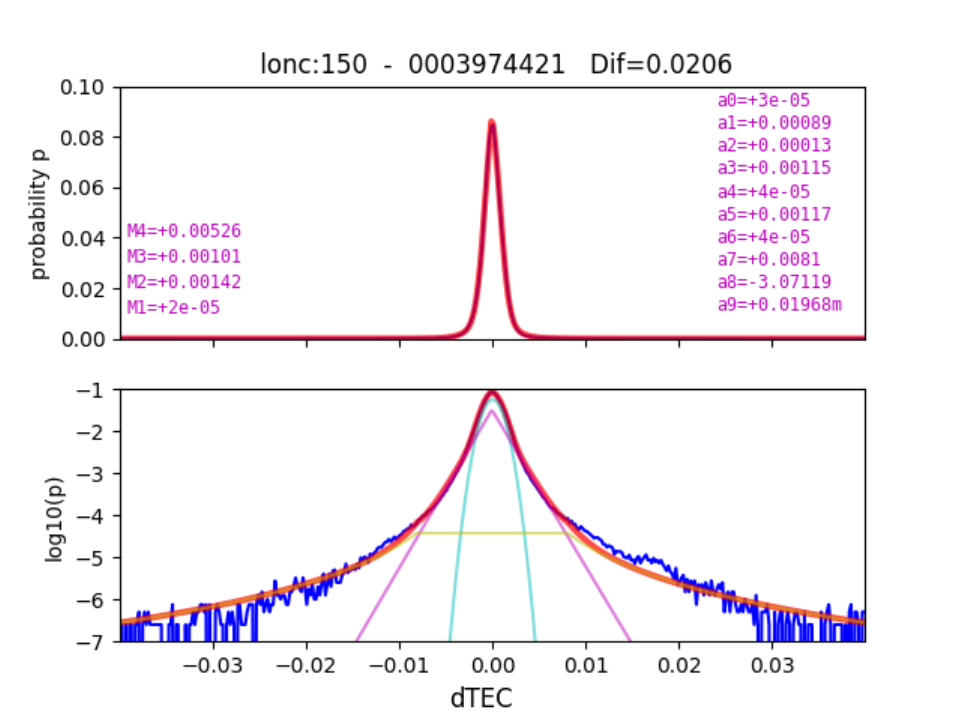}
\includegraphics[width=0.9800\columnwidth]{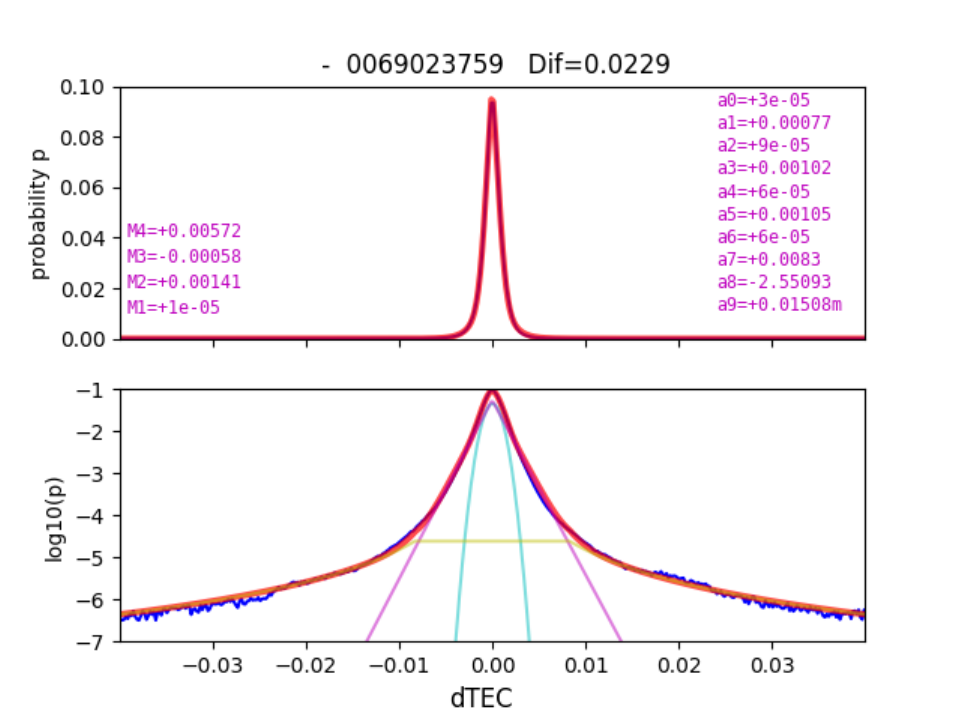}
\caption{Fits to data slice with well over a million dTEC values,
 which clearly indicate the persistence of the 
 core G2E shape, 
 but now also feature well-sampled power law tails.
Line colours \& etc are the same as for Fig. \ref{fig-histo1E2}.
}
\label{fig-histo1E6}
\end{figure}

Particularly at low dTEC event counts, 
 the dTEC histograms can show a wide variety of features, 
 not all of which are captured by our standard G2E fitting function.
However, 
 as the sample counts increased, 
 the histograms do tend to match the G2E form ever more closely, 
 although unexpected features are still possible.
Here we show a range of dTEC histograms
 for increasing event counts
 achieved by choosing data slices with increasing data aggregation.
These automatically generated diagnostics,
 shown here as Figs.  \ref{fig-histo1E2}, \ref{fig-histo1E4}, \ref{fig-histo1E6},
 are shown here to indicate the general nature of the distributions fitted, 
 and not because they contain specific and interesting results.


In particular, 
 Fig. \ref{fig-histo1E2} shows results for slices populated with only hundreds
 of dTEC events/values, 
 and these only hint at the variety of time, space, and band-dependent 
 distributions that exist within the whole dataset.
Note that although they are indeed noisy
 and erratically sampled, 
 the distributions nevertheless can have their own distinct characters 
 according to the contingencies involved in generating the dTEC values
 that go into their shape.
As we move to higher event counts, 
 as indicated e.g. on Fig. \ref{fig-histo1E4}, 
 we see the variety averaged out, 
 albeit still with the G2E fitting model still performing sucessfully 
 even into the poorly sampled wings of the distributions.

Further, 
 note that some distributions have a core dominated by the Gaussian component, 
 but others are instead dominated by the exponential parts; 
 at low numbers of dTEC events (e.g. $n<1000$) there is considerable variation.

As already described, 
 when the data is not divided finely, 
 so that slices might e.g. combining all lpair data, or all lonc data, 
 this means that event counts can easily exceed 10 or 100 thousand, 
 and indeed reach up to approximately 50 million for a whole day.
Consequently 
 the wider tails of the distributions became significantly sampled,
 and the discussed power-law fall-off is revealed,
 as seen on Fig. \ref{fig-histo1E6}. 
However, the G2E fitting typically remained effective --
 albeit not perfect --
 at matching the dominant core and wings of the dTEC distribution.

\begin{figure}
\includegraphics[width=0.9800\columnwidth]{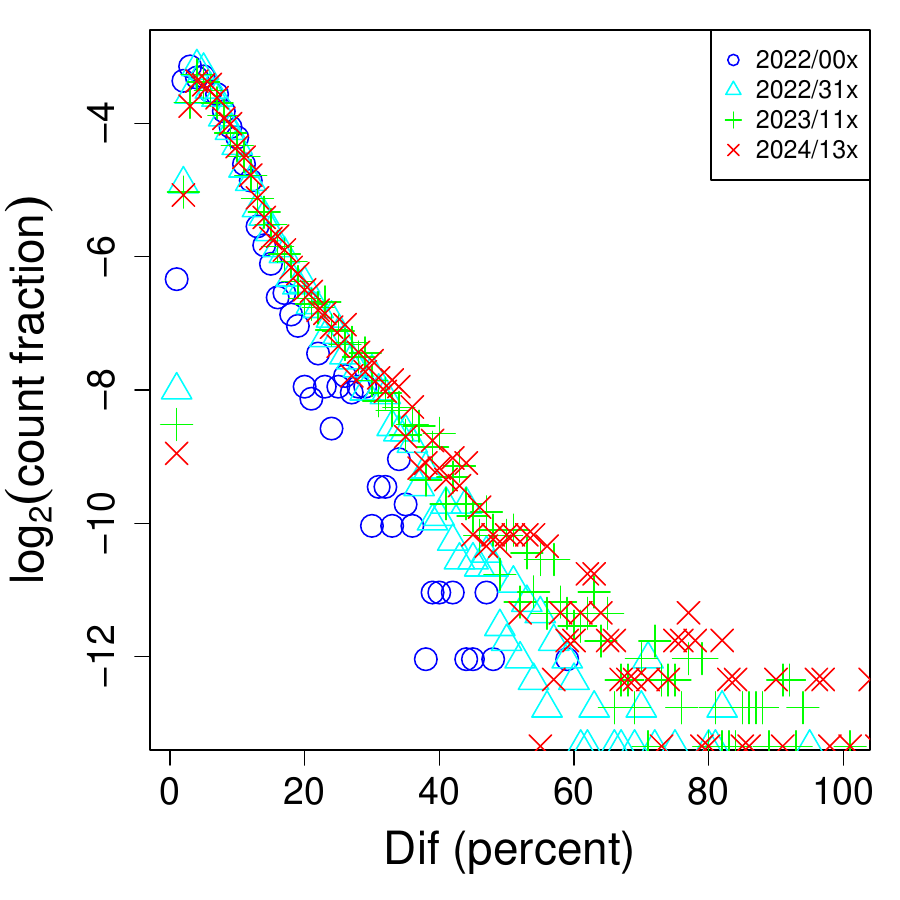}
\caption{Distributions of fitting error estimates 
 for our four sample time intevals. We can see that 
 almost all fits have differences clustered below 10\%, 
 and whilst very poor fits certainly occur, 
 they are relaively rare.
}
\label{fig-dif-cfs}
\end{figure}

In Fig. \ref{fig-dif-cfs} we show the relative likelihoods of 
 fitting errors over our four sample time intervals:
 2022/006 and 007, 
 2022/310 to 319, 
 2023/110 to 119, 
 and 2024/130 to 134.
This error is given as a ``Dif'' value, 
 which is a percentage error derived from the sum of the 
 absolute differences between the data and its fitted counterpart.
However, 
 this difference is only an indirect
 indication of the difference between the fitted width
 and some notional ``true'' width.
Furthermore, 
 not that the error values accumulated in these graphs 
 are for all the slices that are routinely fitted, 
 and many of these share some fraction of the available dTEC events.

\clearpage



\end{document}